\documentclass[a4paper,fleqn]{cas-sc}

\usepackage[authoryear,longnamesfirst]{natbib}
\usepackage{setspace}
\usepackage[linesnumbered,ruled,vlined]{algorithm2e}
\usepackage{graphicx}%
\usepackage{multirow}%
\usepackage{amsmath,amssymb,amsfonts}%
\usepackage{amsthm}%
\usepackage{mathrsfs}%
\usepackage[title]{appendix}%
\usepackage{xcolor}%
\usepackage{textcomp}%
\usepackage{manyfoot}%
\usepackage{booktabs}
\usepackage{multirow}
\usepackage{multirow,booktabs}
\usepackage{makecell}
\usepackage{listings}%
\usepackage{ifthen}
\usepackage{pdflscape}
\usepackage{amssymb}
\usepackage{longtable}
\usepackage{booktabs}
\usepackage{comment}
\usepackage{caption}   
\usepackage{subcaption}
\usepackage{hyperref}
\usepackage{float}
\usepackage{caption}

\hypersetup{
	colorlinks=true,
	linkcolor=red,
	citecolor=blue,
}

\def\tsc#1{\csdef{#1}{\textsc{\lowercase{#1}}\xspace}}
\tsc{WGM}
\tsc{QE}
\tsc{EP}
\tsc{PMS}
\tsc{BEC}
\tsc{DE}

\begin{document}
\let\WriteBookmarks\relax
\def\floatpagepagefraction{1}
\def\textpagefraction{.001}
\shortauthors{Liu et~al.}

\title [mode = title]{Close-enough general routing problem for multiple unmanned aerial vehicles in monitoring missions}                      

\author[nuist]{Huan Liu}
\ead{huanliu@nuist.edu.cn}

\author[polyt,cirrelt]{Michel Gendreau}
\ead{michel.gendreau@polymtl.ca}

\author[csu_jt]{Binjie Xu}
\ead{binjie_xu@csu.edu.cn}

\author[csu_zdh]{Guohua Wu}\cormark[1]
\ead{guohuawu@csu.edu.cn}
\cortext[cor1]{Corresponding author: guohuawu@csu.edu.cn}

\author[csu_jt]{Yi Gu}
\ead{yi.gu@csu.edu.cn}

\affiliation[nuist]{
	organization={School of Automation, Nanjing University of Information Science and Technology},
	addressline={Nanjing}, 
	city={Nanjing},
	postcode={210044}, 
	state={Jiangsu},
	country={China}
}

\affiliation[polyt]{
	organization={Department of Mathematics and Industrial Engineering, Polytechnique Montreal},
	addressline={Montreal}, 
	city={Montreal},
	postcode={H3C 3A7}, 
	state={Quebec},
	country={Canada}
}

\affiliation[cirrelt]{
	organization={Interuniversity Research Centre on Enterprise Networks, Logistics and Transportation (CIRRELT)},
	addressline={Montreal}, 
	city={Montreal},
	postcode={H3C 3A7}, 
	state={Quebec},
	country={Canada}
}

\affiliation[csu_jt]{
	organization={School of Traffic and Transportation Engineering, Central South University},
	addressline={Changsha}, 
	city={Changsha},
	postcode={410075}, 
	state={Hunan},
	country={China}
}

\affiliation[csu_zdh]{
	organization={School of Automation, Central South University},
	addressline={Changsha}, 
	city={Changsha},
	postcode={410083}, 
	state={Hunan},
	country={China}
}

\begin{abstract}
In this paper, we introduce a close-enough multi-UAV general routing problem (CEMUAVGRP) where a fleet of homogeneous UAVs conduct monitoring tasks containing nodes, each of which has its disk neighborhood, and edges, aiming to minimize the total distance. A two-phase iterative method is proposed, partitioning the CEMUAVGRP into a general routing phase where a satisfactory route including required nodes and edges for each UAV is obtained without considering the disk neighborhoods of required nodes, and a close-enough routing phase where representative points are optimized for each required node in the determined route. To be specific, a variable neighborhood descent (VND) heuristic is proposed for the general routing phase, while a second-order cone programming (SOCP) procedure is applied in the close-enough routing phase. These two phases are performed in an iterative fashion under the framework of an adaptive iterated local search (AILS) algorithm until the predefined termination criteria are satisfied.  Extensive experiments and comparative studies are conducted, demonstrating the efficiency of the proposed AILS-VND-SOCP
algorithm and the superiority of disk neighborhoods.
\end{abstract}


\maketitle

\fontsize{11pt}{16.5pt}\selectfont
\onehalfspacing

\section{Introduction}
\label{sec1}
Unmanned aerial vehicles (UAVs), also known as drones, are aircraft that operate without a human pilot onboard. UAVs can be remotely controlled by a human operator or operate autonomously. With the continuous advancements of UAV technology, their applications have expanded beyond the military, finding growing use in various civilian fields as well, such as package delivery \citep{mulumba2024optimization}, maritime pollution detection \citep{liu2023exact}, traffic monitoring \citep{li2018unmanned}, infrastructure inspection \citep{pan2024unmanned}, mapping \citep{chen2021adaptive, glock2020mission}, etc. Among these, UAV monitoring is one of the most valuable emerging applications. Compared to manual methods, it offers a safer, more efficient, and cost-effective solution, particularly for monitoring complex subjects in real-time environments, such as traffic flow and searching for survivors.

In this paper, we focus on monitoring static targets rather than dynamic ones. Mathematically, these targets can be categorized into nodes, edges, and areas. When the UAV monitors an area, it typically follows predetermined edges that effectively cover the region \citep{cao2022concentrated}. Consequently, monitored targets are primarily divided into nodes and edges for analysis. Although an edge consists of multiple nodes, different choices of entry and exit points can significantly impact the route's distance, which is the difference between node routing and arc routing problems.

Numerous studies have focused on node routing and arc routing problems for UAV monitoring. Fang et al. \citep{fang2023routing} investigated a multi-UAV routing problem for monitoring scattered landslide-prone areas, formulating it as a team orienteering problem with mandatory visits. They developed a large neighborhood search algorithm, incorporating a neural network heuristic to solve it. Zhen et al. \citep{zhen2019vehicle} studied a vehicle routing problem for UAV monitoring, where the monitored targets have varying accuracy requirements based on flight altitude. A tabu search was designed to address this issue. Amorosi et al. \citep{amorosi2023multiple} proposed a vehicle-assisted multi-UAV coordinated monitoring framework, in which a single vehicle is equipped with multiple UAVs to monitor arcs. They formulated this problem as a mixed-integer second-order cone programming model aiming to minimize the overall travel time, and proposed both an exact mathematical programming solution and a matheuristic approach. 

Under complicated scenarios, diverse monitored targets often arise. For instance, battlefield missions may involve inspecting specific points, patrolling along lines, and mapping designated areas \citep{atencia2019weighted}. In traffic monitoring, it is crucial to concentrate on high-priority areas such as critical roads, accident-prone intersections, and busy stations.  Compared to monitoring homogeneous targets, integrating heterogeneous targets (such as nodes and edges) into a single route can lead to considerable improvements on the total travel distance \citep{campbell2023multi}. How to determine the sequence of targets involving nodes and edges performed by UAVs, can be defined as multi-UAV general routing problems, which are essential for many UAV monitoring applications. Nevertheless, there is very limited research on this problem.

Moreover, UAVs equipped with sensors can detect targets within a certain range rather than precise positioning. In other words, a target can be identified as long as it falls within the sensing range of airborne sensors. This capability allows UAVs to shorten their total travel distance when visiting node targets. This application can be formulated as classical close-enough traveling salesman problems (CETSP). 

However, this advantage becomes less effective when UAVs are required to monitor edge targets. This is because edge targets (e.g., power lines, pipelines, and boundaries) require specific observation angles to capture valid information. For instance, during power line inspection, UAVs must capture insulators, conductors, and fittings from appropriate angles, while thermal sensors require specific viewpoints to accurately detect hotspots \citep{li2022power,guan2021uav}. 

In contrast, detecting node targets within a certain range has not yet been incorporated into multi-UAV general routing problems.

To address this limitation, we introduce a close-enough multi-UAV general routing problem and develop a two-phase iterative method to solve it. Our main contributions are threefold:

\begin{itemize}
	
	\item To the best of our knowledge, we are the first to address a close-enough multi-UAV general routing problem (CEMUAVGRP) where a fleet of homogeneous UAVs conduct monitoring tasks containing nodes, each of which has its disk neighborhood of radius $r$, and edges. A nested formulation is proposed, where the outer layer is a nonlinear programming for selecting representative points of the required nodes, and the inner layer is an integer linear programming for the general routing problem to minimize the total flight distance considering the maximum flight range of UAVs.
	
	\item We propose a two-phase iterative method, partitioning the CEMUAVGRP into a general routing phase where a satisfactory route including required nodes and edges for each UAV is obtained without considering the disk neighborhoods of required nodes, and a close-enough routing phase where representative points are optimized for each required node in the determined route. To be specific, a variable neighborhood descent (VND) heuristic is developed in the general routing phase, while a second-order cone programming (SOCP) procedure is applied in the close-enough routing phase. These two phases are performed in an iterative fashion under the framework of an adaptive iterated local search (AILS) algorithm until the predefined termination criteria are satisfied. 
	
	\item Extensive experiments on the CEMUAVGRP benchmark instances without disk neighborhoods \citep{campbell2023multi} demonstrate the strong optimization performance of AILS-VND-SOCP. For all 150 benchmark instances, the best solutions obtained by AILS-VND-SOCP show a gap of no more than 2.5\% compared to the best-known solutions, and notably outperform the branch and cut (B\&C) method on 13 instances. Moreover, incorporating disk neighborhoods (i.e., a coverage monitoring strategy) effectively reduces the total flight distance. Increasing the disk neighborhood radius not only shortens the UAVs’ routes but also decreases the number of used UAVs.

\end{itemize}

The remainder of this paper is organized as follows. In Section \ref{sec2}, we conduct a review of existing relevant literature. In Section \ref{sec3}, we formally describe the CEMUAVGRP and derive a nested formulation. In Section \ref{sec4}, we describe the proposed AILS-VND-SOCP in detail. Computational resulted are reported in Section \ref{sec5}. Finally, conclusions are drawn in Section \ref{sec6}.

\section{Literature review}
\label{sec2}
Routing problems with UAVs have recently become a hot topic in the academic community. Traditionally, ground-based vehicles are restricted to existing infrastructure, requiring them to travel along designated roads to reach their destinations. In contrast, UAVs can follow more direct routes, potentially reducing travel costs significantly. However, this flexibility also increases the problem's complexity. For example, an UAV can enter an edge at any point (even intermedidate points), traverse and service part of it, and exit at another point. This capability allows for shorter routes than traditional vehicles, while it transforms the routing problem into a continuous optimization challenge, making it more difficult to solve \citep{campbell2018drone, campbell2023multi}.

The most prevalent UAV routing problems are node routing problems, in which a fleet of UAVs must conduct tasks in each node, with routes starting and ending at a depot. Xia et al. \citep{xia2019drone} were the first to address UAV scheduling for monitoring vessels. They developed a time-expanded network model based on a mixed-integer linear programming (MILP) formulation. To solve this model, they proposed a Lagrangian relaxation-based approach. For an instance with 100 vessels and 3 base stations, the Lagrangian relaxation-based method could obtain desirable solutions with a gap of about 7\% from the best-known solutions within 7200 seconds. Liu et al. \citep{Liu9296564} formulated the multi-UAV task scheduling problem for traffic monitoring as a vehicle routing problem with time windows, aiming to minimize the total distance traveled. They designed a divide-and-conquer framework, consisting of a task allocation phase where tasks are assigned to each UAV without a fixed sequence, and a single-UAV scheduling phase, which generates a feasible route for each UAV. Based on this framework, they proposed a two-phase iterative method to effectively solve the problem. Moreover, to address the computational challenges posed by large-scale problems, Mao et al. \citep{mao2024dl} proposed a double-level deep reinforcement learning (DRL) approach. In this framework, an encoder-decoder structured policy network in the upper-level DRL allocates tasks among UAVs, while an attention-based policy network in the lower-level DRL establishes a route for each UAV. The generalization performance of this double-level DRL was validated on large-scale instances with up to 1500 tasks.

In addition, due to the restricted battery capacity, a vehicle-aided multi-UAV coordinated surveillance mode is investigated. Zeng et al. \citep{zeng2022nested} regarded a truck as the platform for releasing, recovering and recharging UAVs. They developed a mixed integer quadratically constrained programming model to minimize the makespan and proposed a neighborhood search-based heuristic, whose effectiveness was evaluated and demonstrated via a series of comparison experiments with Gurobi and existing heuristics.

The UAV arc routing problem was first introduced by Campbell et al. \citep{campbell2018drone}, where each long arc could be divided into several segments, with a single UAV traversing these segments multiple times to minimize the total duration. Given the limited flight range of each UAV, multiple UAVs were employed  to traverse several arcs, with each arc potentially being serviced by different UAVs. Xu et al. \citep{xu2023gv} introduced a ground-vehicle-assisted multi-UAV arc routing problem, where the ground vehicle provides launch, recovery and recharging support for the UAVs. A mixed integer nonlinear programming formulation was established, and a metaheuristic was designed to solve it.

The UAV general routing problem is associated with planning a minimum-length route starting and ending at the depot, that visits a subset of vertices and traverses a subset of edges and arcs within a given weighted mixed graph \citep{dell2016adaptive}. It was proven to be NP-hard \citep{lenstra1976general}. To the best of our knowledge, Luo et al. \citep{luo2019traffic} first introduced an UAV general routing problem, where multiple UAVs, which are released and recovered by a ground-based vehicle, monitor some edges and nodes in the road-map. The vehicle performs monitoring tasks as well. A MILP was constructed, aiming to minimize the total travel time of the ground-based vehicle and the drones. To tackle the problem efficiently, a two-stage heuristic was proposed, decomposing the original complicated problem into a traveling salesman problem and a multi-UAV routing problem. More specifically, the first stage consists in generating an optimal route for the vehicle completing all tasks, while the second stage assigns some tasks to each UAV from the vehicle considering platform capacity constraints. Its effectiveness was verified by experiments on the Sioux Falls network with 24 nodes and 38 edges. 

Campbell et al. \citep{campbell2023multi} presented the multi-purpose K-drones general routing problem (MP K-DGRP), where each drone can both complete delivery tasks and area-sensing tasks.  They formulated an integer programming model aimed at minimizing the total travel duration for completing both node and edge tasks. A B\&C method was proposed, integrating several families of valid inequalities to enhance solution quality. To further reduce travel duration, they also explored edge splitting, allowing multiple drones to jointly cover an edge by entering and exiting at any point along it. A matheuristic method was developed, which first identifies a set of desirable routes without edge splitting, and then incrementally adds intermediate points on required edges based on these routes. The computational results demonstrated the superiority of multi-purpose drones and of splitting required edges in terms of lowering the travel duration. The proposed B\&C effectively provided optimal solutions for small- and medium-scale instances and generated strong upper bounds for large-scale instances.

Based on the aforementioned MP K-DGRP, Corberán et al. \citep{corberan2024multidepot} extended the problem by introducing multiple depots, defining it as the multidepot drone general routing problem (MDdGRP). Compared to the single-depot version, the MDdGRP allows for more cost-effective routes by incorporating depot selection into the routing process. They established an integer programming formulation and proposed an efficient B\&C involving effective valid inequalities. Two matheuristics were proposed, each involving a general routing phase and an intermediate point selection phase. In the first matheuristic, intermediate points are initially added to required edges before solving the MDdGRP, with additional points incorporated during the problem-solving process. In contrast, the second matheuristic solves the MDdGRP without initially splitting edges, with intermediate points being added only afterward. Computational results demonstrated the superiority of the second approach over the first.

Plana et al. \citep{plana4819017load} further investigated the MP K-DGRP by considering the total weight of the drone, which affects both edge traversal times and delivery times. Specifically, a lighter drone results in shorter edge traversal and delivery times. Since the weight of a drone depends on the cargo it carries, it is crucial to determine the delivery sequence. The authors constructed a mathematical model and developed B\&C with effective valid inequalities, demonstrating strong optimization performance even for large-scale instances involving up to 94 required edges and 7 required nodes.

Based on the above analysis, it is evident that research on the general UAV routing problem is still in its infancy. Most existing studies focus on edge splitting, enabling UAVs to enter and exit edges at any point to minimize travel time. In practice, however, nodes can be detected by UAVs as long as they fall within the field of view of onboard sensors. A node target can be monitored once the UAV reaches any point within its neighborhood. Nevertheless, such considerations have not yet been incorporated into studies on the general routing problem.

These problems are related to the classical close-enough traveling salesman problem (CETSP), which consists in determining the shortest tour that starts and ends at a depot while visiting a set of node targets, each associated with a predefined neighborhood (typically a circular area around the node) \citep{gulczynski2006close, gendreau1997covering}. A node target is considered “visited” when any point within its neighborhood is reached. This feature transforms the problem from a discrete to a continuous optimization problem, thereby significantly increasing its computational complexity. Exact algorithms, like branch and bound \citep{coutinho2016branch}, SOCP \citep{mennell2009heuristics}, etc., can produce optimal solutions but are limited to small-scale instances. To efficiently address the large-scale problems, various heuristics have been developed, including a fast three-step heuristic based on the variable
neighborhood search \citep{wang2019steiner}, as well as a genetic algorithm \citep{di2022genetic} and a memetic algorithm \citep{lei2024effective}.

It is worth noting that some edge targets can only be detected within a specific viewing angle. Jiang et al. \citep{jiang2017uav} highlighted that multi-directional oblique imaging (typically four oblique directions plus one nadir view) can effectively capture the lateral surfaces of insulators, the sag and spacing of conductors, and structural details such as bolts, joints, or corrosion on tower components. El-Sayed et al. \citep{el2023railway} demonstrated that integrating oblique UAV imagery enables better identification of rail fastening systems and sidewall damages, which are difficult to observe from vertical imagery alone. Therefore, in this study, we focus on detecting node targets within a certain sensing range rather than edge targets.

\section{Problem statement}\label{sec3}

\subsection{CEMUAVGRP} \label{sec3.1}

\begin{figure}[ht]
	\centering
	\begin{subfigure}[t]{0.5\linewidth}
		\captionsetup{justification=centering} 
		\includegraphics[width=\linewidth]{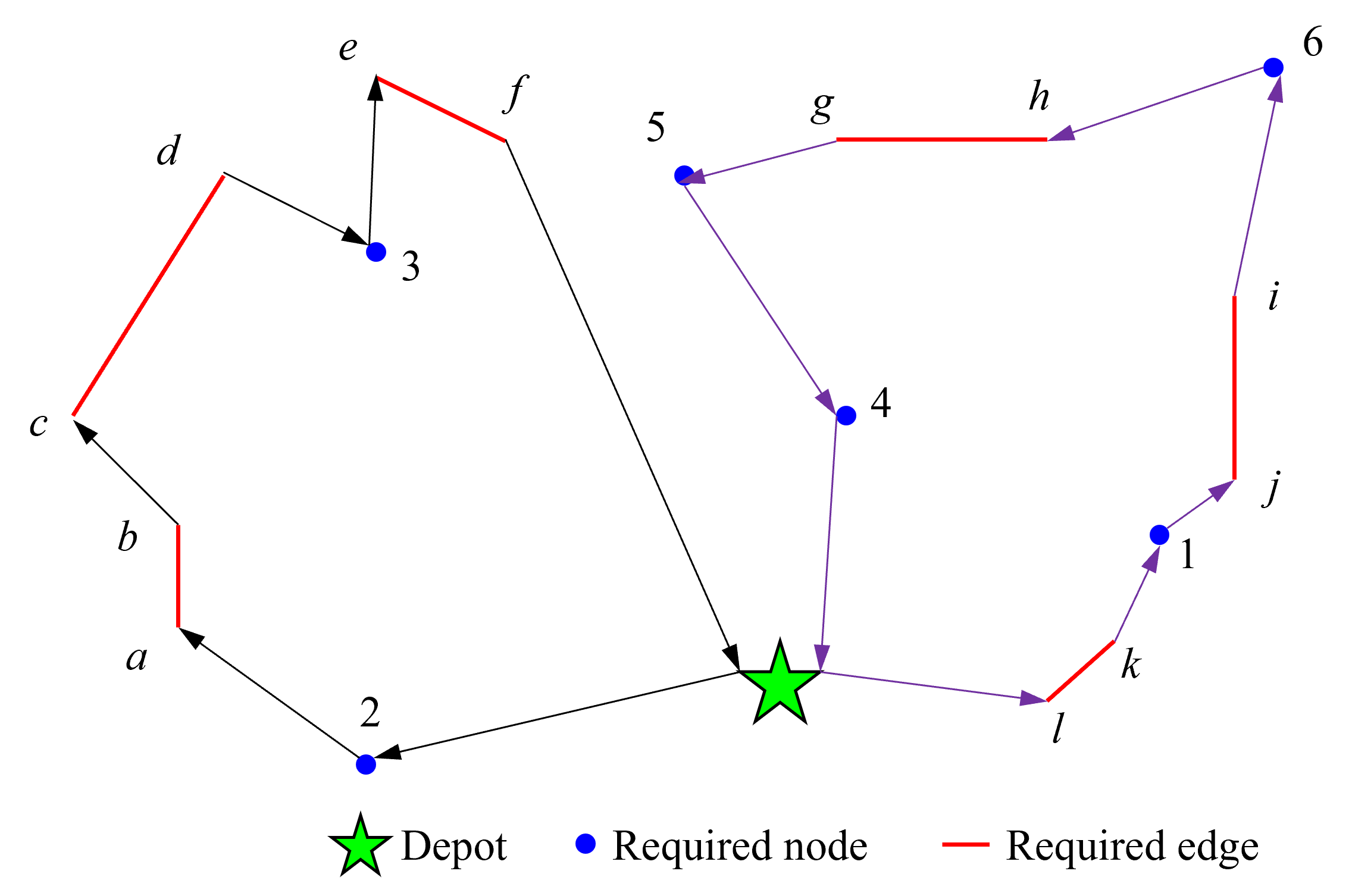}
		\caption{Without the coverage monitoring strategy}
	\end{subfigure}
	\vspace{3mm} 
	
	\begin{subfigure}[t]{0.5\linewidth}
		\captionsetup{justification=centering} 
		\includegraphics[width=\linewidth]{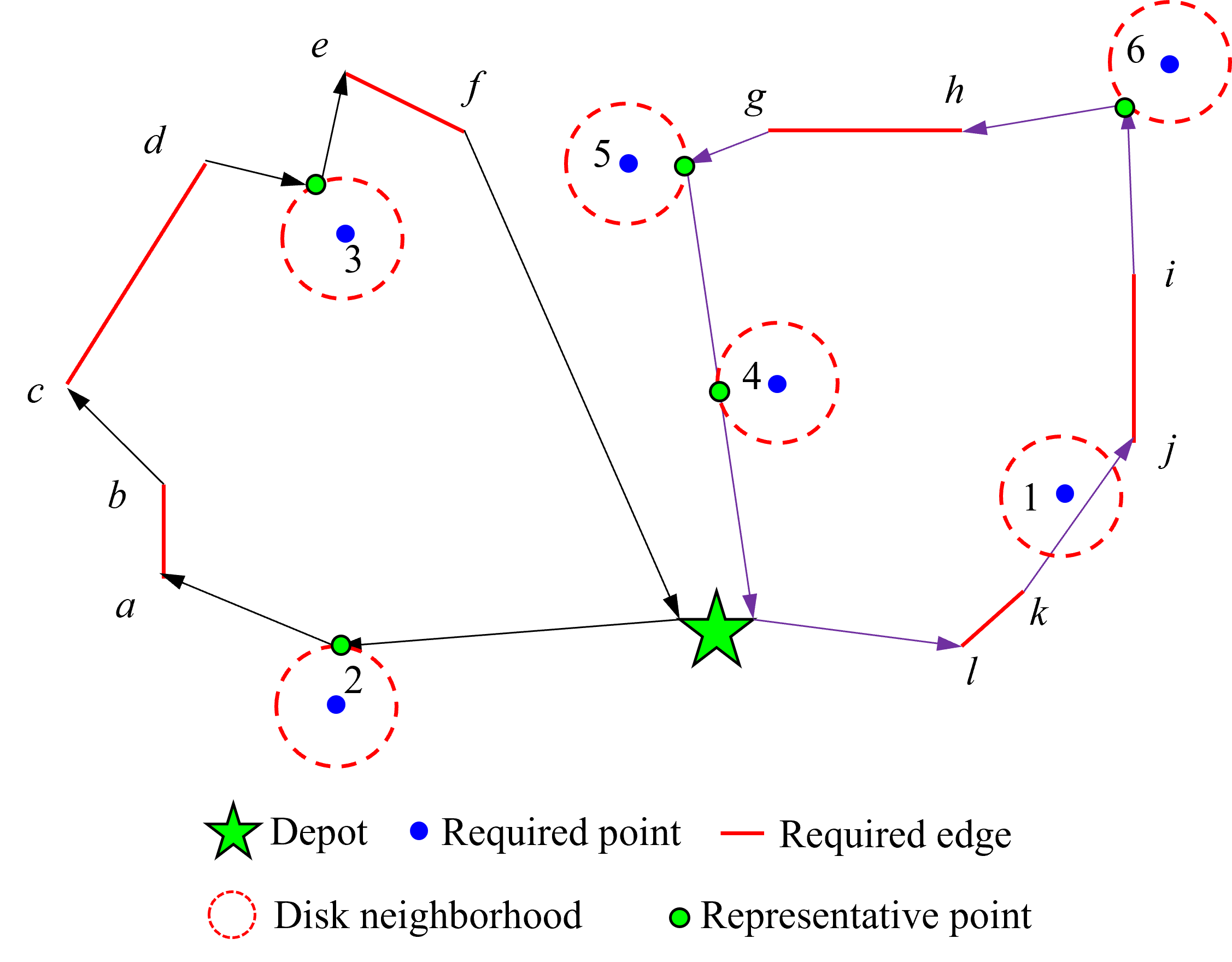}
		\caption{With the coverage monitoring strategy}
	\end{subfigure}
	
	\vspace{3mm}
	\caption{CEMUAVGRP}
	\label{fig1}
\end{figure}

In the CEMUAVGRP, each UAV departs from the depot and returns to it after completing its assigned tasks, which include both required nodes and edges. The objective is to minimize the total flight distance. To better reflect the actual monitoring scenario, two key factors are considered in this problem: (1) Required edges: In addition to required nodes, required edges are also considered, whose entry and exit directions have an impact on the UAV's flight distance. (2) A coverage monitoring strategy: The airborne sensor can monitor required nodes within a specified range. As illustrated in Fig.~\ref{fig1}(b), each required node is associated with a disk neighborhood (a red dashed circle). A node is considered monitored once the UAV traverses its corresponding neighborhood, which contributes to a further reduction in the flight distance. For instance, as shown in Fig.~\ref{fig1}(a), after monitoring the required edges $(l,k)$, the UAV must deliberately fly to required node \# 1 to conduct monitoring, resulting in additional flight distance. However, in Fig.~\ref{fig1}(b), when the coverage monitoring strategy is adopted, the UAV passes through the disk neighborhood of node \# 1 while flying toward the edge $(j, i)$ after monitoring $(l,k)$. Consequently, the node is monitored without requiring any additional detour.

More specifically, this problem can be decomposed into two sub-problems: one is the general routing sub-problem, which ignores the coverage monitoring strategy. As shown in Fig.~\ref{fig1}(a), each required node is associated with a fixed representative point, and the goal is to generate a route for each UAV. The other is the close-enough routing sub-problem. As shown in Fig.~\ref{fig1}(b), this sub-problem focuses on selecting the optimal representative point in the disk neighborhood of each required node, while keeping the task sequence in the existing route unchanged. 

The first sub-problem focuses on route optimization at the macro level, aiming to improve the overall route structure. This involves adjusting the task sequence within a single route and exchanging tasks between different routes, belonging to discrete optimization problems. In contrast, the second sub-problem addresses route optimization at the micro level, where the objective is to optimize the representative point for each required node while keeping the task sequence in the existing route unchanged. This corresponds to continuous optimization problems.

\subsection{Formulation} \label{sec3.2}

Since CEMUAVGRP can be divided into a general routing sub-problem and a close-enough routing sub-problem, a nested formulation is proposed as follows:

(1) The outer layer consists in selecting the representative points $\mathbf{p}=(p_i^x, p_i^y)$ within the disk neighborhood $\mathcal{D}_i$ for each required node $i \in V_R$, where $V_R$ denotes the set of required nodes. Each required node $i \in V_R$ has a disk neighborhood $\mathcal{D}_i$ of radius $r_i$, and its coordinates are $\mathbf{c}_i = (c_i^x, c_i^y)$. A required node is considered visited if the UAV reaches any point within its corresponding neighborhood. Accordingly, the constraint requires that each representative point lies within its corresponding neighborhood:
\begin{equation}
	(p_i^x-c_i^x)^2 + (p_i^y-c_i^y)^2 \leq r_i^2,\forall i \in V_R. \label{Eq:eq1}
\end{equation}

\begin{table}[h]
	\caption{Notations} \label{tab1}
	\footnotesize
	\renewcommand{\arraystretch}{1}
	\begin{tabular}{p{0.2\textwidth}p{0.7\textwidth}}
		\toprule
		Notations & Description \\
		\midrule
		\multicolumn{2}{c}{\textit{Sets}} \\
		\midrule
		$ U $ & a set of UAVs \\
		$ V_R $ & a set of required nodes\\
		$ V_E $ & a set of points incident with required edges \\
		$ V_c $ &  a set of initial “must be visited” nodes, $V_c=\{0\} \cup V_R \cup V_E$  \\
		$ G $ & an undirected graph $G=(V_p,E)$, where $V_p=\{0\}\cup V(\textbf{p})\cup V_E, E = \{(i,j)\subseteq V_p \times V_p:i<j \}$. $\textbf{p}$ denotes a vector of representative points for point $i\in V_c$, and $V(\textbf{p})$ is the set of “must be visited” nodes. 0 denotes the depot.\\
		$ E_R $ & a set of required edges, $E_R=\{e_1,e_2, \dots, e_n \},e_q=(i,j),i,j\in V_E, i<j$ \\
		$ E_{NR} $ & a set of non-required edges \\
		\midrule
		\multicolumn{2}{c}{\textit{Parameters}} \\
		\midrule
		$ k $ & the index of UAVs \\
		$r_i$ &  the radius of the disk neighborhood of the point $i\in V_c$, $r_i=0$ if $i \in V_c\backslash V_R$\\
		$\textbf{c} = (c_i^x,c_i^y)$ &  the coordinates of the point $i\in V_c$  \\
		$ d_{i,j} $ & the travel distance between point $i$ and point $j$ \\
		$ L_k $ & the flight range of UAV $k$ \\
		\midrule
		\multicolumn{2}{c}{\textit{Decision variables}} \\
		\midrule
		$ x_{i,j}^k $ & a binary value, equals 1 if UAV $k$ traverses the edge and travels  from point $i$ to point $j$, $(i,j)\in E$ \\
		$ z_{i}^k $ & a binary value, equals 1 if UAV $k$ services required node $i\in V_R$ \\
		$\textbf{p}=(p_i^x,p_i^y)$ & continuous values, the coordinates of a representative point in the disk neighborhood $\mathcal{D}_i$ (within radius $r_i$) of the point $ i\in V_c$ \\ 
		\bottomrule
	\end{tabular}
	
\end{table}

(2) The inner layer is a general routing problem defined on the undirected graph $G=(V_p,E)$, where $V_p=\{0\}\cup V(\textbf{p})\cup V_E, E = \{(i,j)\subseteq V_p \times V_p:i<j\}$. In this formulation, $\textbf{p}$ denotes a vector of representative points, and $V(\textbf{p})$ is the set of “must be visited” nodes. 0 denotes the depot. $V_E$ denotes the set of points incident to required edges. $V_c$ denotes the set of initial “must be visited” nodes, including 0, $V_E$ and $V_R$. $E_R$ and $E_{NR}$ denote the sets of required and non-required edges, respectively. The required edge $ (i,j) $ is regarded as completed if the UAV traverses either from $ i $ to $ j $ or from $ j $ to $ i $.

There is a fleet of $k_{max}$ UAVs, i.e., $U$. The flight range of UAV $k$ is $L_k$. The objective is to generate $k_{max}$ routes with minimum total flight distance, starting and ending at the depot, that jointly traverse all the required edges and visit the required nodes exactly once, so that the flight distance of each route does not exceed the flight range.

Three families of decision variables are defined: (1) $x_{i,j}^k$ represents whether UAV $k$ traverses the edge $(i,j) \in E$. If UAV $k$ flies from $i$ to $j$, $x_{i,j}^k$ is equal to 1; otherwise, $x_{i,j}^k$ is equal to 0. (2) A binary variable $z_i^k$ takes the value 1 if the point $i$ is served by UAV $k$ and 0 otherwise. (3) $\textbf{p}=(p_i^x,p_i^y)$ denotes the coordinates of the representative point in the disk neighborhood $\mathcal{D}_i$ (within radius $r_i$) of the point $ i\in V_c$. The notations are listed in Table~\ref{tab1}.

Given a subset $S\subseteq V_p, \delta(S)=(S:V_p \backslash S)$ denotes the edge set with one endpoint in $S$ and the other in $V_p \backslash S$. $E(S)=(S:S)$ denotes the edge set with both endpoints in $S$. For any subset $F\subseteq E$, we denote $F_R=F \cap E_R,F_{NR}=F \cap E_{NR}$. $F_R$ represents the required edges in $F$, and $F_{NR}$ represents the non-required edges in $F$.

The inner problem can be formulated as follows. 
\begin{equation}
	\min f=\sum \limits _{k\in U} \sum \limits _{(i,j)\in E} d_{i,j}(x_{i,j}^{k}+x_{j,i}^{k}). \label{Eq:eq2}
\end{equation}
s.t.
\begin{equation}
	\sum \limits _{j:(0,j)\in \delta(0)} x_{0,j}^{k} = 1, \forall k \in U, \label{Eq:eq3}
\end{equation}
\begin{equation}
	\sum \limits _{j:(j,0)\in \delta(0)} x_{j, 0}^{k} = 1, \forall k \in U, \label{Eq:eq4}
\end{equation}
\begin{equation}
	\sum \limits _{k\in U} (x_{i,j}^{k} + x_{j,i}^{k})= 1, \forall (i,j) \in E_R, \label{Eq:eq5}
\end{equation}
\begin{equation}
	\sum \limits _{k\in U}z_i^k= 1, \forall i \in V_R, \label{Eq:eq6}
\end{equation}
\begin{equation}
	\sum \limits _{j:(i,j) \in \delta(i)} x_{i,j}^{k}= z_i^k, \forall i \in V_R, \forall k \in U, \label{Eq:eq7}
\end{equation}
\begin{equation}
	\sum \limits _{j:(i,j)\in \delta (i)} x_{i,j}^{k} - \sum \limits _{j:(j,i)\in \delta (i)} x_{j,i}^{k}=0,\forall i \in \{0\} \cup V_E, \forall k \in U, \label{Eq:eq8}
\end{equation}
\begin{equation}
\sum \limits _{(i,j)\in \delta_{R}(S)} (x_{i,j}^{k} + x_{j,i}^{k}) + \sum \limits _{(i,j)\in \delta_{NR}(S)} (x_{i,j}^{k} + x_{j,i}^{k})\geq 2(x_{u,v}^k+x_{v,u}^k), \forall S \subset V_p \backslash \{0\}, \forall (u,v) \in E(S), \forall k \in U, \label{Eq:eq9}
\end{equation}
\begin{equation}
	d_{i,j}^2 = (p_i^x-p_j^x)^2+(p_i^y-p_j^y)^2,\forall i,j \in V_c, \label{Eq:eq10}
\end{equation}
\begin{equation}
	p_i^x=c_i^x, p_i^y=c_i^y,\forall i \in \{0\} \cup V_E, \label{Eq:eq11}
\end{equation}
\begin{equation}
	\sum \limits _{(i,j)\in E} (x_{i,j}^{k}+x_{j,i}^{k})d_{i,j} - L_k \leq 0,\forall k \in U, \label{Eq:eq12}
\end{equation}
\begin{equation}
	x_{i,j}^{k} \in \{0,1\},\forall (i,j) \in E, \forall k \in U, \label{Eq:eq13}
\end{equation}
\begin{equation}
	z_{i}^{k} \in \{0,1\},\forall i \in V_R, \forall k \in U, \label{Eq:eq14}
\end{equation}
where the objective function is to minimize the total travel distance of all UAVs when completing all tasks. Constraints (\ref{Eq:eq3}) and (\ref{Eq:eq4}) mandate that each UAV must start from and return to the depot. Constraint (\ref{Eq:eq5}) mean that each required edge must be serviced by only one UAV. Constraints (\ref{Eq:eq6})-(\ref{Eq:eq7}) restrict that each required node must be serviced by only one UAV. Constraint (\ref{Eq:eq8}) is a flow balance that the in-degree of each non-required node is equal to its out-degree. Constraint (\ref{Eq:eq9}) is connectivity inequalities. Due to the degree conditions, the UAV must cross any given cutset an even number of times. To be more specific, if the set $S$ includes the required tasks (edges or nodes), the UAV must go in and then out of the set $S$. Otherwise, the number of crossing the cutset for an UAV is equal to zero. This constraint can avoid sub-tours and also ensures that each single tour is connected to the depot. Constraint (\ref{Eq:eq10}) computes the distance between two points. Constraint (\ref{Eq:eq11}) enforces that representative points corresponding to the required edges and the depot coincide with their original locations. Constraint (\ref{Eq:eq12}) restricts the maximum travel distance for each UAV. Constraints (\ref{Eq:eq13})-(\ref{Eq:eq14}) limit the value range for each decision variable.

\section{Solution methodology}
\label{sec4}

The close-enough multi-UAV general routing problem can be divided into two independent sub-problems: one is a general routing problem that each route of an UAV can be generated, which connects nodes and edges without the consideration of disk neighborhoods. The other is a close-enough routing problem that the best representative point in each disk of required nodes is selected while the sequence of edges and points in one route is fixed. Although these two sub-problems can be solved independently in theory, they are in fact highly interdependent. In the general routing sub-problem, any changes to the route structure, such as the addition or removal of nodes, the adjustment of edges, or alterations in the visiting sequence, will directly impact the selection of representative points in the close-enough routing problem. In contrast, the optimal representative points generated in the close-enough routing problem may trigger further adjustments to the route structure. Thus, while logically distinct, the two sub-problems are tightly coupled and mutually constrained during the solution process.

\begin{figure}[ht]%
	\centering
	\includegraphics[width=0.8\textwidth]{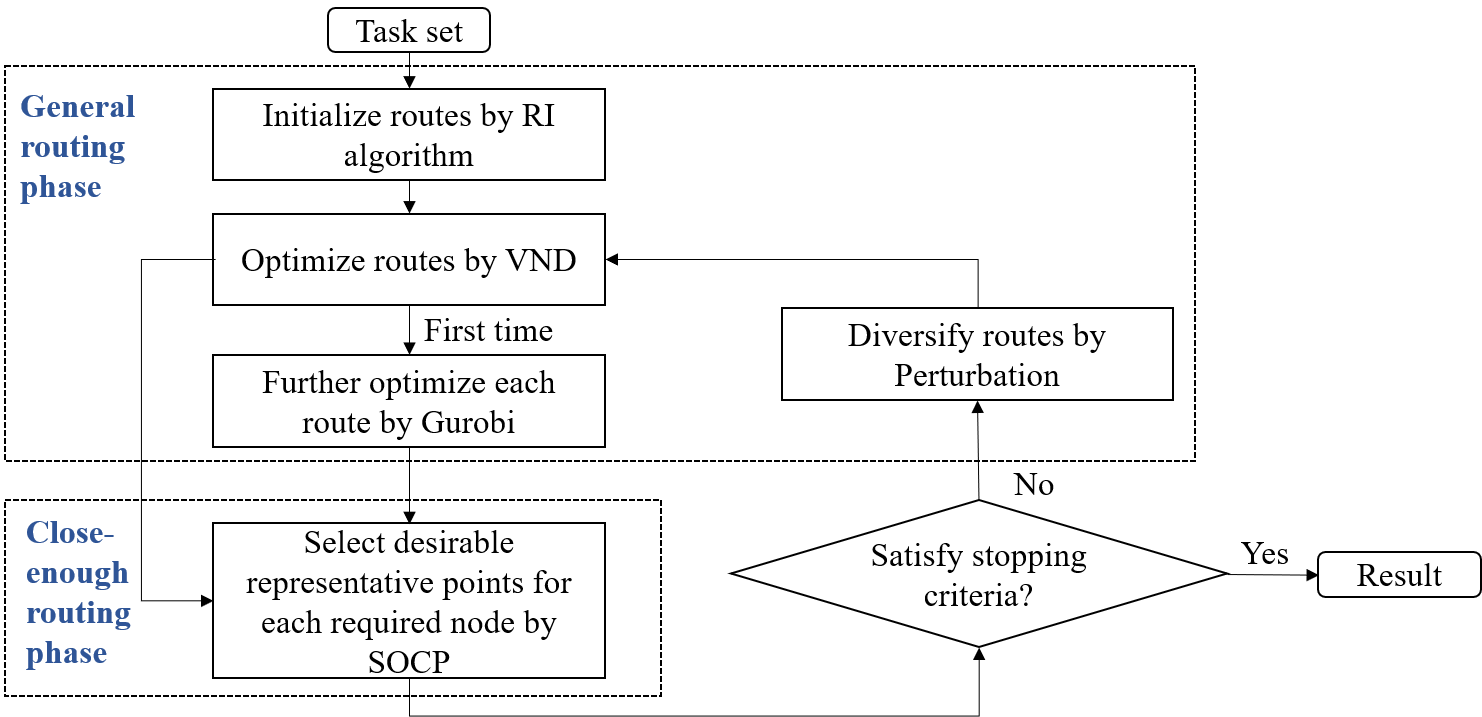}
	\caption{The two-phase iterative optimization framework}\label{fig2}
\end{figure}

This means that the problem division may have an impact on the global optimality of the original problem. Thus, the problem division operator and the sub-problem solving procedure can be conducted iteratively and interactively to alleviate the negative impact of the problem division. Based on it, we propose a two-phase iterative optimization framework involving a general routing phase and a close-enough routing phase, as shown in Fig.~\ref{fig2}.

In the general routing phase, we propose a regret insertion algorithm (RI) to generate an initial solution, where the UAV must travel to the exact location of each required node. This initial solution is then optimized using VND, incorporating both intra-route and inter-route moves. Each route is further optimized by the commercial solver Gurobi, but only for the initial solution. In the close-enough routing phase, we apply SOCP to select the satisfactory representative point for each required node, while maintaining a fixed task sequence in the route. The two phases are performed iteratively and interactively within the framework of AILS until the predefined stopping criteria are met. The pseudo-code of the proposed AILS-VND-SOCP is depicted in Algorithm \ref{Alg:AILS-VND-SOCP}.

\begin{algorithm}
	\caption{AILS-VND-SOCP}\label{Alg:AILS-VND-SOCP}
	\KwIn{maximum number of iterations $MaxIt$, maximum number of consecutive unimproved solutions $it_{max}$, minimum number of tasks destroyed $\tau_{min}$, maximum number of tasks destroyed $\tau_{max}$, maximum number of consecutive unimproved solutions with the same number of tasks destroyed $L_{max}$, reset-to-best trigger interval $\theta$, execution frequency of increasing the acceptance threshold $\beta$; }
	\KwOut{$ P_{best} $;}
	Initialize: $unimproved \leftarrow 0,rejected\leftarrow 0,improved \leftarrow 0,noim \leftarrow 0, \alpha_1 \leftarrow 1,\alpha_2 \leftarrow 1,\eta \leftarrow 1+\alpha_1*\alpha_2, \tau \leftarrow \tau_{min}$\;
	$S_0\leftarrow RegretInsertion()$ /*\textit{Initial solution construction: Algorithm} \ref{Alg:RI}*/\;
	$S_1\leftarrow VND(S_0)$/*\textit{Local search: Algorithm} \ref{Alg:VND}*/\;
	$S\leftarrow Gurobi(S_1)$\;
	$P\leftarrow SOCP(S)$/*\textit{Representative point selection}*/\;
	$S_{best}\leftarrow S, P_{best} \leftarrow P$\;
	\For { $i \leftarrow 1:MaxIt$}
	{
		$S^{'}\leftarrow Perturbation(S,\tau)$/*\textit{Perturbation: Algorithm} \ref{Alg:Per}*/\;
		$S_{new}\leftarrow VND(S^{'})$/*\textit{Local search: Algorithm} \ref{Alg:VND}*/\;
		$P_{new}\leftarrow SOCP(S_{new})$/*\textit{Representative point selection}*/\;
		\If{$f(P_{new}) < f(P_{best})$}
		{
			$P_{best}\leftarrow P_{new},S_{best}\leftarrow S_{new}$\;
			$unimproved\leftarrow 0, rejected\leftarrow 0, improved\leftarrow improved + 1$,
			$noim\leftarrow 0$\;
		}
		\Else
		{
			$unimproved\leftarrow unimproved+ 1,rejected\leftarrow rejected + 1,noim\leftarrow noim + 1$\;
		}
		
		\If{$ f(P_{new}) \leq \eta*f(P_{best}) $}
		{
			$ P \leftarrow P_{new}, S \leftarrow S_{new} $\;
			$ \alpha_1 \leftarrow f(P_{best})/f(S_0), \alpha_2 \leftarrow improved/i, \eta \leftarrow 1+\alpha_1*\alpha_2 $\;
		}	
		\If{$rejected \ \% \ \beta = 0$}
		{
			$ \eta \leftarrow \eta+\alpha_1*\alpha_2, rejected \leftarrow 0$\;
		}
		\If{$ noim \geq L_{max}$}
		{
			$ \tau \leftarrow min(\tau+1, \tau_{max}), noim \leftarrow 0  $\;         
		}
		\ElseIf{$ noim = 0$}
		{
			$ \tau \leftarrow \tau_{min} $\;
		}
		\If{$unimproved \ \% \ \theta =0$}
		{
			$ P \leftarrow P_{best}, S \leftarrow S_{best} $\;
		}
		\If{$unimproved = it_{max}$}
		{
			\textbf{break}
		}
		
	}	          
\end{algorithm}
The main procedures of AILS-VND-SOCP are as follows:

\textit{Step 1 (Parameter initialization):} Some parameters are initialized, including the counts of consecutive unimproved, rejected and improved solutions, the number of consecutive unimproved solutions with the same number of tasks destroyed, the parameters related to the threshold acceptance and the number of tasks destroyed (Line 1).

\textit{Step 2 (Initial solution construction):} Several initial solutions are generated by RI, where the best is selected as $S_0$ (Line 2).

\textit{Step 3 (General routing):} The initial solution is optimized by VND to reduce the total distance (Line 3). To further minimize the total distance, Gurobi is applied for additional optimization (Line 4). 

\textit{Step 4 (Close-enough routing):} Given that the order of tasks in each route remains fixed, SOCP is then used to optimize representative points for required nodes (Line 5), and the best-so-far solutions with and without the consideration of disk neighborhoods, i.e., $S_{best}$ and $P_{best}$, are defined (Line 6).

\textit{Step 5 (Iterative optimization):} The solution is optimized using iterative local search (ILS), which sequentially applies the perturbation mechanism, VND, SOCP and threshold acceptance procedures until the predefined stopping criteria are met (Lines 7-28). Two stopping criteria are defined, either of which can terminate the algorithm: reaching a maximum number of iterations or encountering a maximum number of consecutive unimproved solutions. If the new solution is better than the current best-known solution, it is designated as the best-so-far solution, and the counts for unimproved, rejected, and improved solutions are updated accordingly (Lines 11-13). Otherwise, the counts for unimproved solutions, rejected solutions, and special unimproved solutions are each incremented by 1 (Lines 14-15). The new solution is accepted as the incumbent solution if its distance is less than $\eta$ times that of the best-so-far distance (Lines 16-17). In this situation, the threshold parameters $ \alpha_1, \alpha_2 $ and threshold value $\eta$ are adjusted (Line 18). To escape from local optima, $\eta$ is increased to accept more poor solutions when the number of rejected solutions reaches a specified limit (Lines 19-20). Meanwhile, the number of tasks destroyed in the perturbation procedure is incremented by 1 but always remains within the range $[\tau_{min}, \tau_{max}]$ (Lines 21-24). The incumbent solution is set as the best-so-far solution when it cannot be further improved in successive iterations, ensuring that the search process does not deviate from the best-so-far solution (Lines 25-26).

\subsection{Initial solution construction}
\label{subsec4.1}

The initial solution plays a crucial role in achieving high-quality results when exploring the search space using metaheuristics. Starting the search from a point close to the optimal solution is expected to reduce the algorithm's computation time. To this end, we propose a RI heuristic, incorporating look-ahead information into the greedy algorithm to improve its mytopic nature \citep{ropke2006adaptive, wang2020iterative}. 

Given a task $i\in T$ and the $k$-th route $\mathcal{R}_k$, $k \in U$, the insertion cost for adding task $i$ into the end of route $\mathcal{R}_k$ is denoted as $\delta_i^{k}$. Furthermore, $\delta_i^{c_p}$ represents the $p$-th lowest insertion cost when inserting task $i$ into the end of route $\mathcal{R}_{c_p}, c_p \in U$. The regret value can be defined as the difference between the cost of inserting the task $i$ into the best and second-best routes, i.e., $RV_i = \delta_i^{c_2} - \delta_i^{c_1}$. Then, the task $i$ with the highest regret value, i.e., $ argmax_{i}\ RV_i$, will be inserted into its best route. If more than one task shares the largest regret value, one task is selected randomly. This process is repeated until all tasks are inserted into the routes. Due to the efficiency of RI, we can generate a high-quality initial solution by selecting the best one from the set of solutions. The pseudo-code of RI is presented in Algorithm \ref{Alg:RI}.

\begin{algorithm}
	\caption{RI}\label{Alg:RI}
	\KwIn{Task set $T$, UAV set $U$, maximum number of iterations $\rho$; }
	\KwOut{$ S_0$ with $\lvert U \rvert$ routes;}
	Initialize $\lvert U \rvert$ empty routes, i.e., $S_0$\;
	\For{$it\leftarrow1:\rho$}
	{
		$cad \leftarrow T$\;
		\While{$cad \neq \emptyset$}
		{
			Compute the regret value $RV_i$ for each task $i \in cad$\;
			Select the task with the largest regret value\;
			Insert the task into its best route\;
			Remove the task from $cad$\;
		}
		Obtain a feasible solution $S$\;
		\If{$f(S)<f(S_0)$}
		{
			$S_0 \leftarrow S$\;
		}        
	}   
	
\end{algorithm}

\subsection{Local search}
\label{subsec4.2}

In AILS, local search  is responsible for intensification, aiming to search intensively promising parts of search space, which can enhance the possibility of finding the optimal solution. VND has proven highly effective in exploring promising parts \citep{hansen2019variable, ferreira2024variable} and is therefore employed in this process. 

Starting from a solution $x$, VND sequentially explores the $\mu$-th neighborhood $\mathcal{N}_\mu(x)$  to find a better solution. If there is a better solution $x'$, $x'$ is accepted as the new solution, and the search restarts from the first neighborhood structure. If no improvement is found, the search proceeds to the $\mu$+1-th neighborhood. The optimization process terminates once no further improvements can be found. The pseudo-code of VND is given in Algorithm \ref{Alg:VND}.

\begin{algorithm}
	\caption{VND}\label{Alg:VND}
	\KwIn{Initial solution $S$, maximum iterations $l_{max}$, neighborhood structures; }
	\KwOut{$S_{b}$;}
	$S_{b} \leftarrow S, \mu \leftarrow 1$\;
	\For{$l\leftarrow 1:l_{max}$}
	{
		$S' \leftarrow FindBestSolution( \mathcal{N}_\mu(S) )$\;
		\If{$f(S')<f(S_{b})$}
		{
			$S \leftarrow S'$\;
			$S_{b} \leftarrow S'$\;
			$\mu \leftarrow 1$\;
		}
		\Else
		{
			$\mu \leftarrow \mu+1$\;
		}
	}   
	
\end{algorithm}

In VND, sophisticated neighborhood structures are essential for effective search. Each structure is designed to effectively explore promising areas of the search space. By applying them sequentially, the algorithm can quickly identify high-potential areas and then conduct an intensive search within those areas to converge to an optimal solution. Five neighborhood structures are designed as follows:

\textit{(1) 2-opt Operator $\mathcal{N}_1$}: Exchange two edges (excluding the required edges) in the route with two other edges. 

\textit{(2) Flip Operator $\mathcal{N}_2$}: Reverse the direction of one required edge to lower the total distance. This procedure is repeated until an improvement occurs or until the direction of all required edges is reversed without any improvement.

\textit{(3) Destroy-Repair Operator $\mathcal{N}_3$}: Randomly choose $\zeta$ tasks from their corresponding routes and then relocate them one by one by \textit{Regret-Repair}. This procedure is repeated until one improved solution is found or until $\xi$ consecutive iterations without any improvement are performed.

\textit{Regret-Repair}: Compute the regret value $rv_i$ for each task $i$, defined as the difference between its lowest insertion cost and second-lowest insertion cost, i.e., $rv_i = \delta_i^{c_2} - \delta_i^{c_1}$. The task with the highest regret value $ argmax_{i}\ rv_i$ is prioritized and inserted first into its best route. If a task cannot be inserted into any existing route, a new route is created. This process repeats until all removed tasks are successfully inserted into routes.  

\textit{(4) Chain-Insert Operator $\mathcal{N}_4$}: Remove a chain with $\gamma$ consecutive tasks from its current route and reinsert it into another route to minimize the total distance. The process begins with $\gamma=1$. If removing and reinserting all chains with $\gamma$ consecutive tasks fails to reduce the total distance, $\gamma$ is incremented by 1. This process repeats until an improved solution is found or until no further improvement occurs when $\gamma$ reaches $\gamma_{max}$.

\textit{(5) Chain-Exchange Operator $\mathcal{N}_5$}: Exchange a chain of $\gamma_1$ consecutive tasks from one route with a chain of $\gamma_2$ consecutive tasks from another route to minimize the total distance. The process begins with $\gamma_1 =\gamma_2 =1$. If no improved solution can be detected, $\gamma_2$ is incremented by 1. The process is repeated until $\gamma_2$ reaches $\gamma_{max}$ without any improvement. At this point, $\gamma_2$ is reset to 1, and $\gamma_1$ is incremented by 1. The procedure continues until an improvement is achieved or until $\gamma_1 = \gamma_2 = \gamma_{max}$ without any improvement.

\subsection{Perturbation}
\label{subsec4.3}

In AILS,  the perturbation mechanism promotes diversity to prevent the algorithm from becoming trapped in local optima. Within this mechanism, one destroy operator and one repair operator are randomly selected. The number of tasks destroyed, $\tau$, varies with iterations rather than remaining fixed. When AILS gets trapped into local optima for a long time, a high-intensity perturbation is applied. Otherwise, a low-intensity perturbation is expected. To this end, $\tau$ is incremented by 1 (up to $\tau_{max}$) to expand solution search space after $L_{max}$ consecutive iterations without any improvement. If an improved solution is found, $\tau$ is reset to $\tau_{min}$. The pseudo-code of Perturbation is presented in Algorithm \ref{Alg:Per}.

\begin{algorithm}
	\caption{Perturbation}\label{Alg:Per}
	\KwIn{number of tasks destroyed $\tau$, maximum number of iteration $ \lambda$, $S$; }
	\KwOut{$S'$;}
	\For{$i\leftarrow 1:\lambda$}
	{
		$\mathcal{N}_d \leftarrow$ randomly select an operator from $DestroyList$\;
		$\mathcal{N}_r \leftarrow$ randomly select an operator from $RepairList$\;
		$S' \leftarrow $ use $\mathcal{N}_d, \mathcal{N}_r$ on $S$\;
		\If{$S' \neq S$}
		{
			\textbf{break}
		}
	}
	
\end{algorithm}
The effectiveness of the perturbation mechanism hinges on the design of the destroy and repair operators. Four destroy operators are designed, which are as follows:

\textit{(1) Random-Destroy}: Randomly select $\tau$ tasks and remove them from their routes.

\textit{(2) Worst-Destroy}: The removal cost of a task is defined as the difference in costs between before and after removal. The task with the highest removal cost is first removed. The process continues until $\tau$ tasks are all removed. 

\textit{(3) Node-Destroy}: Randomly select $\tau$ node tasks and remove them from their routes, provided that $\tau \leq \lvert V_R \rvert$.

\textit{(4) Edge-Destroy}: Randomly select $\tau$ edge tasks and remove them from their routes, provided that $\tau \leq \lvert E_R \rvert$.

Along with \textit{Regret-Repair} described in Section \ref{subsec4.2}, two additional repair operators are proposed as follows:

\textit{(1) Random-Repair}: Randomly insert each task into a route according to the sequence where they are removed. If the task cannot be inserted successfully, a new route is created. The process continues until the $\tau$ tasks are all reinserted. 

\textit{(2) Greedy-Repair}: The insertion cost is defined as the difference in cost between before and after insertion. The task with the cheapest insertion cost is first inserted into its best route. If the task cannot be inserted successfully, a new route is created. The process continues until the $\tau$ tasks are all reinserted.

\subsection{Acceptance criterion}
\label{subsec4.4}

The acceptance criterion determines whether the solution obtained after the perturbation and local search mechanisms in an iteration becomes the incumbent solution for the next iteration. Two options are available. On the one hand, a new solution can be accepted only when an improvement occurs. On the other hand, a worse solution may be accepted, as exploring the neighborhood of poorer solutions could potentially lead to a better result \citep{kirkpatrick1983optimization, gendreau2010handbook}. 

AILS adopts threshold acceptance, which relies on a threshold value to decide whether a solution of given quality should be accepted. When $f(P_{new}) \leq \eta*f(P_{best})$, the new solution can be accepted as the incumbent solution for the next iteration. To efficiently approach the optimal solution, a threshold value $\eta$ is adjusted according to the current optimization condition instead of remaining fixed. It can be calculated as follows:
\begin{equation}
	\eta = 1+\alpha_1 * \alpha_2, \label{Eq:eq15} 
\end{equation}
\begin{equation}
	\alpha_1 = f(P_{best})/f(S_0), \label{Eq:eq16}
\end{equation}
\begin{equation}
	\alpha_2 = improved/i, \label{Eq:eq17}
\end{equation}
where $P_{best}$ denotes the best-so-far solution considering disk neighborhoods of required nodes. $S_0$ indicates the initial solution without the consideration of these disk neighborhoods. $improved$ represents the number of iterations with improvements, and $i$ is the current iteration. $\alpha_1$ and $\alpha_2$ are set to 1 at the beginning and then updated when a new solution is accepted. Moreover, when the number of rejected solutions reaches a number, $\eta$ is re-increased to accept poorer solutions as follows:
\begin{equation}
	\eta = \eta + \alpha_1 * \alpha_2. \label{Eq:eq18}
\end{equation}

\subsection{Representative point selection}
\label{subsec4.5}

Each required node has a disk neighborhood with a radius of $r$. The required node can be visited when the UAV passes through this neighborhood. Representative points of each required node can be optimized to further reduce the distance of routes, and SOCP is employed for this optimization. Given a route $\mathcal{R}$, $pt_i$ denotes the representative point of vertex $i, i\in \{1, 2, \dots, \lvert \mathcal{R} \rvert \}$, where $pt_1$ and $pt_{\lvert \mathcal{R} \rvert}$ indicate the depot. The location of $pt_i$ is defined by variables $p_i^x$ and $p_i^y$. Let $c_i^x$ and $c_i^y$ be the location of the target covered by representative point $pt_i$. SOCP is then formulated as follows:
\begin{equation}
	min \sum g_i, \label{Eq:eq19}
\end{equation}
\begin{equation}
	f_i = p_i^x -p_{i+1}^x, \forall i \in \{1, 2, \dots, \lvert\mathcal{R}\rvert-1\},\label{Eq:eq20}
\end{equation}
\begin{equation}
	u_i = p_i^y -p_{i+1}^y, \forall i \in \{1, 2, \dots, \lvert\mathcal{R}\rvert-1\},\label{Eq:eq21}
\end{equation}
\begin{equation}
	s_i = c_i^x -p_i^x, \forall i \in \{1, 2, \dots, \lvert\mathcal{R}\rvert\},\label{Eq:eq22}
\end{equation}
\begin{equation}
	h_i = c_i^y -p_i^y, \forall i \in \{1, 2, \dots, \lvert\mathcal{R}\rvert\},\label{Eq:eq23}
\end{equation}
\begin{equation}
	g_i^2 \geq f_i^2 + u_i^2, \forall i \in \{1, 2, \dots, \lvert\mathcal{R}\rvert-1\},\label{Eq:eq24}
\end{equation}
\begin{equation}
	r_i^2 \geq s_i^2 + h_i^2, \forall i \in \{1, 2, \dots, \lvert\mathcal{R}\rvert\},\label{Eq:eq25}
\end{equation}
\begin{equation}
	g_i \geq 0, \forall i \in \{1, 2, \dots, \lvert\mathcal{R}\rvert-1\}.\label{Eq:eq26}
\end{equation}

$g_i$ is defined as the Euclidean distance between representative points $pt_i$ and $pt_{i+1}$. The objective is to minimize the total distance among representative points in the route. The constraints can be categorized into three main types: (1) The route length of $i$-th sub-route ((\ref{Eq:eq20}), (\ref{Eq:eq21}) and (\ref{Eq:eq24})); (2) Each representative point remains within its disk neighborhood $\mathcal{D}_i$ ((\ref{Eq:eq22}), (\ref{Eq:eq23}) and (\ref{Eq:eq25})). If vertex $i$ is a required node, i.e., $i\in V_R$, the radius $r_i$ equals $r$; otherwise, $r_i$ equals zero; (3) The non-negativity of the length of the $i$-th sub-route (\ref{Eq:eq26}).

\section{Computational results}
\label{sec5}

In this section, we assess the performance of our proposed AILS-VND-SOCP method without disk neighborhoods by comparing it to best-known solutions. We also verify the effectiveness of the coverage monitoring strategy in reducing the total distance. The effectiveness of re-increasing the acceptance threshold is then evaluated, followed by a sensitivity analysis on the disk neighborhood radius. All experiments are implemented in Python and run on a PC with an Intel Core i7-12700F 2.10 GHz CPU, 32 GB of RAM, using the Windows 11 operating system. Gurobi 9.5.2 is used to further optimize the solution obtained from VND and to solve the SOCP, with the maximum solving times set to 120 seconds and 200 seconds, respectively.

\subsection{Comparison with best-known solutions}
\label{subsec5.1}

\subsubsection{Benchmark instances}
\label{subsubsec5.1.1}

To the best of our knowledge, there are no open benchmark instances available for CEMUAVGRP. To evaluate the optimization performance of our proposed AILS-VND-SOCP, we first compare it with the best-known solutions without disk neighborhoods of required nodes (i.e., coverage monitoring strategy) generated by B\&C within 7200 seconds \citep{campbell2023multi}. The algorithm without considering the disk neighborhoods is designated as AILS-VND-SOCPI. However, the node tasks in \citep{campbell2023multi} are designed for delivery, with each task having a specific demand. To facilitate comparison, we use only the instances where each required node has a unit demand. This allows us to define the UAV’s capacity as the maximum number of required nodes it can complete.

Two different types of instances, namely Type \text{I} and Type \text{II}, are generated as shown in Appendix A Figs. A.1-A.2. The main difference between them is that Type \text{II} instances have more required edges, which are longer and closer to each other. Specifically, Type I includes 10 basic instances, while Type II includes 20 basic instances, resulting in a total of 30 basic instances, labeled sequentially from C1 to C30 (corresponding to T1\_6\_6\_16\_2u to T2\_10\_10\_16\_2u). Considering the influence of UAV maximum flight range, each basic instance is further divided into five sub-instances, denoted as “Cx-x”. For example, the instance where an UAV with a flight range of 3000 executes T1\_6\_6\_16\_2u is labeled as C1-1. Based on this design, a total of 150 test instances are generated, as summarized in Table~\ref{tab2}.
$N_v$ denotes the number of required nodes. $N_e$ represents the number of required edges. $L$ indicates the maximum flight range of an UAV, and $Q$ refers to the maximum number of required nodes that a single UAV can complete.

{\fontsize{8pt}{7.5pt}\selectfont
	\begin{longtable}{*{12}{c}}
		\caption{Overview of instances} \label{tab2} \\ 
		\toprule
		Instance & $N_v$ & $N_e$ & $L$ & $Q$ & No. & Instance & $N_v$ & $N_e$ & $L$ & $Q$ & No.\\
		
		\hline
		\endfirsthead
		
		\multicolumn{3}{l}{Continued from previous page} \vspace{2mm}\\
		\toprule
		Instance & $N_v$ & $N_e$ & $L$ & $Q$ & No. & Instance & $N_v$ & $N_e$ & $L$ & $Q$ & No.\\
		
		\hline
		\endhead
		
		\hline
		\endfoot
		
		\hline
		\endlastfoot

		\multirow{5}{*}{T1\_6\_6\_16\_2u} & \multirow{5}{*}{6} & \multirow{5}{*}{15}
		& 3000 & 4 & C1-1 & \multirow{5}{*}{T1\_6\_8\_16\_2u} & \multirow{5}{*}{7} & \multirow{5}{*}{21}
		& 4000 & 4 & C2-1 \\
		& & & 2000 & 3 & C1-2 & & & & 3500 & 3 & C2-2 \\
		& & & 1600 & 2 & C1-3 & & & & 3000 & 2 & C2-3 \\
		& & & 1500 & 2 & C1-4 & & & & 2000 & 2 & C2-4 \\
		& & & 1300 & 2 & C1-5 & & & & 1800 & 2 & C2-5 \\
		\cmidrule(lr){2-6}\cmidrule(lr){8-12}
		
		\multirow{5}{*}{T1\_7\_7\_16\_2u} & \multirow{5}{*}{5} & \multirow{5}{*}{14}
		& 4000 & 3 & C3-1 & \multirow{5}{*}{T1\_8\_6\_16\_2u} & \multirow{5}{*}{8} & \multirow{5}{*}{23}
		& 4000 & 5 & C4-1 \\
		& & & 3000 & 2 & C3-2 & & & & 3500 & 3 & C4-2 \\
		& & & 2000 & 2 & C3-3 & & & & 3000 & 2 & C4-3 \\
		& & & 1800 & 2 & C3-4 & & & & 2000 & 2 & C4-4 \\
		& & & 1700 & 1 & C3-5 & & & & 1800 & 2 & C4-5 \\
		\cmidrule(lr){2-6}\cmidrule(lr){8-12}
		
		\multirow{5}{*}{T1\_8\_8\_16\_2u} & \multirow{5}{*}{6} & \multirow{5}{*}{16}
		& 4000 & 4 & C5-1 & \multirow{5}{*}{T1\_8\_10\_16\_2u} & \multirow{5}{*}{10} & \multirow{5}{*}{41}
		& 7000 & 7 & C6-1 \\
		& & & 2700 & 3 & C5-2 & & & & 6000 & 5 & C6-2 \\
		& & & 2200 & 2 & C5-3 & & & & 4000 & 4 & C6-3 \\
		& & & 1900 & 2 & C5-4 & & & & 3500 & 3 & C6-4 \\
		& & & 1800 & 2 & C5-5 & & & & 3000 & 3 & C6-5 \\
		\cmidrule(lr){2-6}\cmidrule(lr){8-12}
		
		\multirow{5}{*}{T1\_9\_9\_16\_2u} & \multirow{5}{*}{10} & \multirow{5}{*}{29}
		& 7000 & 6 & C7-1 & \multirow{5}{*}{T1\_10\_8\_16\_2u} & \multirow{5}{*}{12} & \multirow{5}{*}{36}
		& 7000 & 7 & C8-1 \\
		& & & 6000 & 4 & C7-2 & & & & 5000 & 5 & C8-2 \\
		& & & 3500 & 3 & C7-3 & & & & 3500 & 4 & C8-3 \\
		& & & 2800 & 3 & C7-4 & & & & 3000 & 3 & C8-4 \\
		& & & 2700 & 2 & C7-5 & & & & 2500 & 3 & C8-5 \\
		\cmidrule(lr){2-6}\cmidrule(lr){8-12}
		
		\multirow{5}{*}{T1\_10\_10\_16\_2u} & \multirow{5}{*}{11} & \multirow{5}{*}{27}
		& 7000 & 6 & C9-1 & \multirow{5}{*}{T1\_12\_12\_16\_2u} & \multirow{5}{*}{12} & \multirow{5}{*}{49}
		& 8000 & 7 & C10-1 \\
		& & & 6000 & 4 & C9-2 & & & & 6000 & 5 & C10-2 \\
		& & & 3500 & 3 & C9-3 & & & & 4500 & 4 & C10-3 \\
		& & & 2800 & 3 & C9-4 & & & & 4000 & 3 & C10-4 \\
		& & & 2700 & 2 & C9-5 & & & & 3500 & 3 & C10-5 \\
		\cmidrule(lr){2-6}\cmidrule(lr){8-12}
		
		\multirow{5}{*}{T2\_6\_6\_16\_1u} & \multirow{5}{*}{7} & \multirow{5}{*}{26}
		& 6000 & 4 & C11-1 & \multirow{5}{*}{T2\_6\_6\_16\_2u} & \multirow{5}{*}{13} & \multirow{5}{*}{20}
		& 5000 & 7 & C12-1 \\
		& & & 4000 & 3 & C11-2 & & & & 3100 & 5 & C12-2 \\
		& & & 3500 & 2 & C11-3 & & & & 3000 & 4 & C12-3 \\
		& & & 3000 & 2 & C11-4 & & & & 2500 & 3 & C12-4 \\
		& & & 2500 & 2 & C11-5 & & & & 2000 & 3 & C12-5 \\
		\cmidrule(lr){2-6}\cmidrule(lr){8-12}

		\multirow{5}{*}{T2\_6\_8\_16\_1u} & \multirow{5}{*}{6} & \multirow{5}{*}{38}
		& 7000 & 4 & C13-1 & \multirow{5}{*}{T2\_6\_8\_16\_2u} & \multirow{5}{*}{13} & \multirow{5}{*}{28}
		& 7000 & 7 & C14-1 \\
		& & & 4500 & 3 & C13-2 & & & & 4500 & 5 & C14-2 \\
		& & & 4000 & 2 & C13-3 & & & & 4000 & 4 & C14-3 \\
		& & & 3000 & 2 & C13-4 & & & & 3000 & 3 & C14-4 \\
		& & & 2800 & 2 & C13-5 & & & & 2800 & 3 & C14-5 \\
		\cmidrule(lr){2-6}\cmidrule(lr){8-12}
		
		\multirow{5}{*}{T2\_6\_10\_16\_1u} & \multirow{5}{*}{7} & \multirow{5}{*}{53}
		& 9000 & 4 & C15-1 & \multirow{5}{*}{T2\_6\_10\_16\_2u} & \multirow{5}{*}{12} & \multirow{5}{*}{36}
		& 7200 & 7 & C16-1 \\
		& & & 7500 & 3 & C15-2 & & & & 5200 & 5 & C16-2 \\
		& & & 6000 & 2 & C15-3 & & & & 4200 & 4 & C16-3 \\
		& & & 4500 & 2 & C15-4 & & & & 3500 & 3 & C16-4 \\
		& & & 3500 & 2 & C15-5 & & & & 3300 & 3 & C16-5 \\
		\cmidrule(lr){2-6}\cmidrule(lr){8-12}
		
		\multirow{5}{*}{T2\_7\_7\_16\_1u} & \multirow{5}{*}{8} & \multirow{5}{*}{48}
		& 7500 & 5 & C17-1 & \multirow{5}{*}{T2\_7\_7\_16\_2u} & \multirow{5}{*}{14} & \multirow{5}{*}{28}
		& 5500 & 8 & C18-1 \\
		& & & 6500 & 3 & C17-2 & & & & 4000 & 5 & C18-2 \\
		& & & 5000 & 3 & C17-3 & & & & 3500 & 4 & C18-3 \\
		& & & 4000 & 2 & C17-4 & & & & 2800 & 3 & C18-4 \\
		& & & 3200 & 2 & C17-5 & & & & 2500 & 3 & C18-5 \\
		\cmidrule(lr){2-6}\cmidrule(lr){8-12}
		
		\multirow{5}{*}{T2\_7\_9\_16\_1u} & \multirow{5}{*}{7} & \multirow{5}{*}{63}
		& 9000 & 4 & C19-1 & \multirow{5}{*}{T2\_7\_9\_16\_2u} & \multirow{5}{*}{14} & \multirow{5}{*}{33}
		& 7500 & 8 & C20-1 \\
		& & & 7000 & 3 & C19-2 & & & & 5500 & 5 & C20-2 \\
		& & & 6000 & 3 & C19-3 & & & & 5000 & 4 & C20-3 \\
		& & & 4000 & 2 & C19-4 & & & & 4000 & 3 & C20-4 \\
		& & & 3500 & 2 & C19-5 & & & & 3000 & 3 & C20-5 \\
		\cmidrule(lr){2-6}\cmidrule(lr){8-12}
		
		\multirow{5}{*}{T2\_8\_8\_16\_1u} & \multirow{5}{*}{9} & \multirow{5}{*}{54}
		& 9500 & 5 & C21-1 & \multirow{5}{*}{T2\_8\_8\_16\_2u} & \multirow{5}{*}{16} & \multirow{5}{*}{43}
		& 8000 & 9 & C22-1 \\
		& & & 7000 & 5 & C21-2 & & & & 6500 & 6 & C22-2 \\
		& & & 6000 & 3 & C21-3 & & & & 5500 & 5 & C22-3 \\
		& & & 5000 & 2 & C21-4 & & & & 4000 & 4 & C22-4 \\
		& & & 4000 & 2 & C21-5 & & & & 3500 & 3 & C22-5 \\
		\cmidrule(lr){2-6}\cmidrule(lr){8-12}
		
		\multirow{5}{*}{T2\_8\_10\_16\_1u} & \multirow{5}{*}{9} & \multirow{5}{*}{70}
		& 11000 & 5 & C23-1 & \multirow{5}{*}{T2\_8\_10\_16\_2u} & \multirow{5}{*}{16} & \multirow{5}{*}{40}
		& 7600 & 9 & C24-1 \\
		& & & 8000 & 4 & C23-2 & & & & 6000 & 6 & C24-2 \\
		& & & 6500 & 3 & C23-3 & & & & 4500 & 5 & C24-3 \\
		& & & 6000 & 3 & C23-4 & & & & 4000 & 4 & C24-4 \\
		& & & 4500 & 2 & C23-5 & & & & 3400 & 3 & C24-5 \\
		\cmidrule(lr){2-6}\cmidrule(lr){8-12}
		
		\multirow{5}{*}{T2\_9\_9\_16\_1u} & \multirow{5}{*}{10} & \multirow{5}{*}{73}
		& 11000 & 6 & C25-1 & 	\multirow{5}{*}{T2\_9\_9\_16\_2u} & \multirow{5}{*}{19} & \multirow{5}{*}{27}
		& 7000 & 10 & C26-1 \\
		& & & 8000 & 4 & C25-2 & & & & 5000 & 7 & C26-2 \\
		& & & 6500 & 3 & C25-3 & & & & 4000 & 5 & C26-3 \\
		& & & 5000 & 2 & C25-4 & & & & 3500 & 4 & C26-4 \\
		& & & 4500 & 2 & C25-5 & & & & 3100 & 4 & C26-5 \\
		\cmidrule(lr){2-6}\cmidrule(lr){8-12}
		
		\multirow{5}{*}{T2\_9\_10\_16\_1u} & \multirow{5}{*}{10} & \multirow{5}{*}{76}
		& 12000 & 6 & C27-1 & \multirow{5}{*}{T2\_9\_10\_16\_2u} & \multirow{5}{*}{19} & \multirow{5}{*}{37}
		& 10000 & 10 & C28-1 \\
		& & & 10000 & 4 & C27-2 & & & & 8000 & 7 & C28-2 \\
		& & & 7500 & 3 & C27-3 & & & & 5500 & 5 & C28-3 \\
		& & & 6000 & 2 & C27-4 & & & & 4000 & 4 & C28-4 \\
		& & & 5000 & 2 & C27-5 & & & & 3500 & 4 & C28-5 \\
		\cmidrule(lr){2-6}\cmidrule(lr){8-12}
		
		\multirow{5}{*}{T2\_10\_10\_16\_1u} & \multirow{5}{*}{10} & \multirow{5}{*}{84}
		& 14000 & 6 & C29-1 & \multirow{5}{*}{T2\_10\_10\_16\_2u} & \multirow{5}{*}{20} & \multirow{5}{*}{61}
		& 14000 & 11 & C30-1 \\
		& & & 12000 & 4 & C29-2 & & & & 12000 & 7 & C30-2 \\
		& & & 8000 & 3 & C29-3 & & & & 8000 & 6 & C30-3 \\
		& & & 6000 & 2 & C29-4 & & & & 6000 & 5 & C30-4 \\
		& & & 5000 & 2 & C29-5 & & & & 5000 & 4 & C30-5 \\
		
		\hline
	\end{longtable}
}

\subsubsection{Parameter setting}
\label{subsubsec5.1.2}
An iterative trial-and-error approach is adopted to determine the best parameters. The relevant parameters are summarized in Table~\ref{tab3}.

\begin{table}[htp]
	\caption{Parameter setting} \label{tab3}
	\footnotesize
	\centering
	\renewcommand{\arraystretch}{1.2}
	\begin{tabular}{p{2cm}|p{8cm}|p{2cm}}
		\toprule
		Parameters & Description & Value \\
		\midrule
		$ Maxit $ & The maximum number of iterations & 100 \\
		$ it_{max} $ & The maximum number of consecutive unimproved solutions & 30 \\
		$ \rho $ & The number of initial solutions & 10 \\
		$ \tau_{min} $ & The minimum number of tasks destroyed in Perturbation mechanism & 3 \\
		$ \tau_{max} $ & The maximum number of tasks destroyed in Perturbation mechanism & 8 \\ 
		$ L_{max} $ & The maximum number of consecutive unimproved solutions with the same number of tasks destroyed in Perturbation mechanism & 8 \\
		$ \lambda $ & The maximum number of iterations in Perturbation mechanism & 5 \\                
		$ \beta $ & The parameter related to increasing the acceptance threshold & 10 \\
		$ timeg_{max} $ & The time limitation of Gurobi & 120 seconds \\
		$ times_{max} $ & The time limitation of SOCP & 200 seconds \\
		$ \zeta $ & The number of tasks destroyed in VND & [2, 8] \\
		$ \xi $ & The maximum number of iterations in \textit{Destroy-Repair Operator} & 5 \\
		$ \gamma_{max} $ & The maximum length of a chain in \textit{Chain-Insert Operator} and \textit{Chain-Exchange Operator} & 3 \\
		\bottomrule
	\end{tabular}
\end{table}

\subsubsection{Experiment results}
\label{subsubsec5.1.3}

\begin{figure}[htb]
	\centering
	\begin{subfigure}[t]{0.7\linewidth}
		\captionsetup{justification=centering} 
		\includegraphics[width=\linewidth]{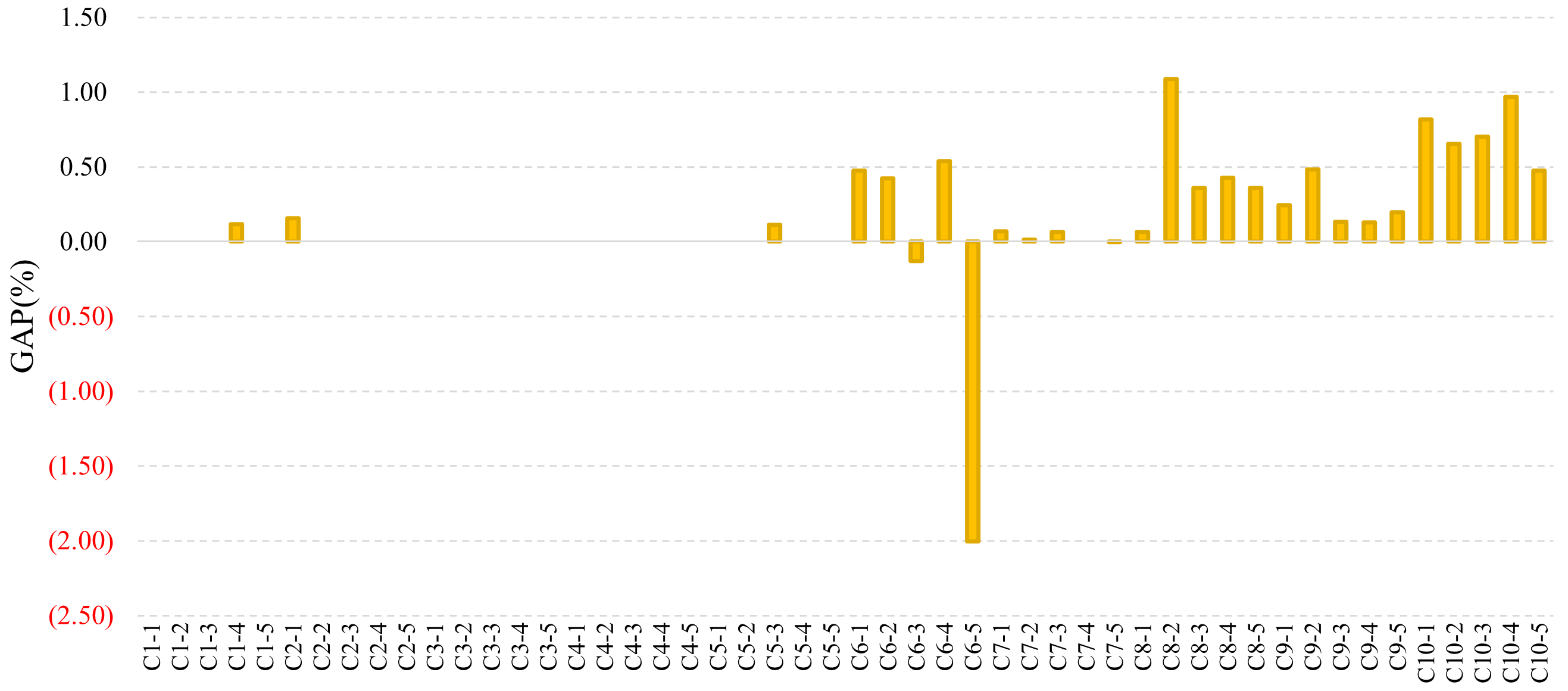}
		\caption{Type \text{I} }
	\end{subfigure}
	\vspace{3mm} 
	
	\begin{subfigure}[t]{0.7\linewidth}
		\captionsetup{justification=centering} 
		\includegraphics[width=\linewidth]{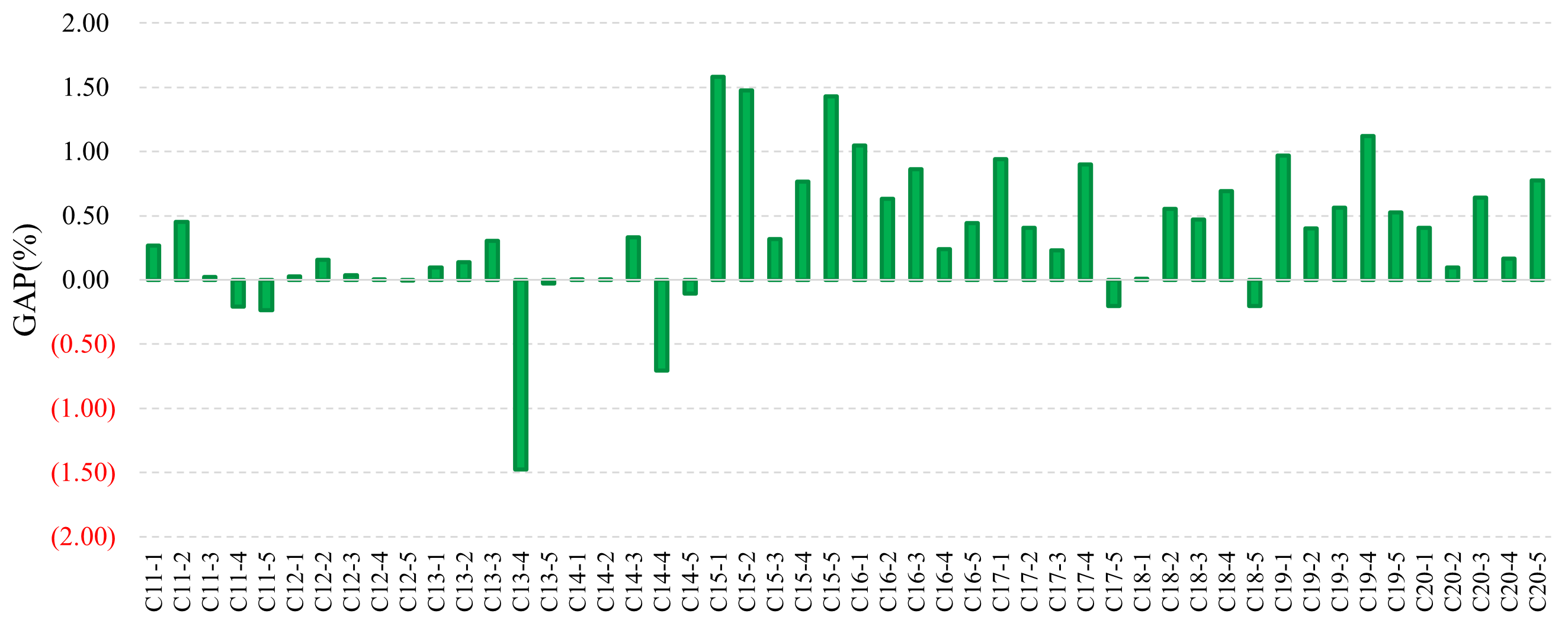}
		\caption{Type \text{II}: The first 50 instances}
	\end{subfigure}
	\vspace{3mm} 
	
	\begin{subfigure}[t]{0.7\linewidth}
		\captionsetup{justification=centering} 
		\includegraphics[width=\linewidth]{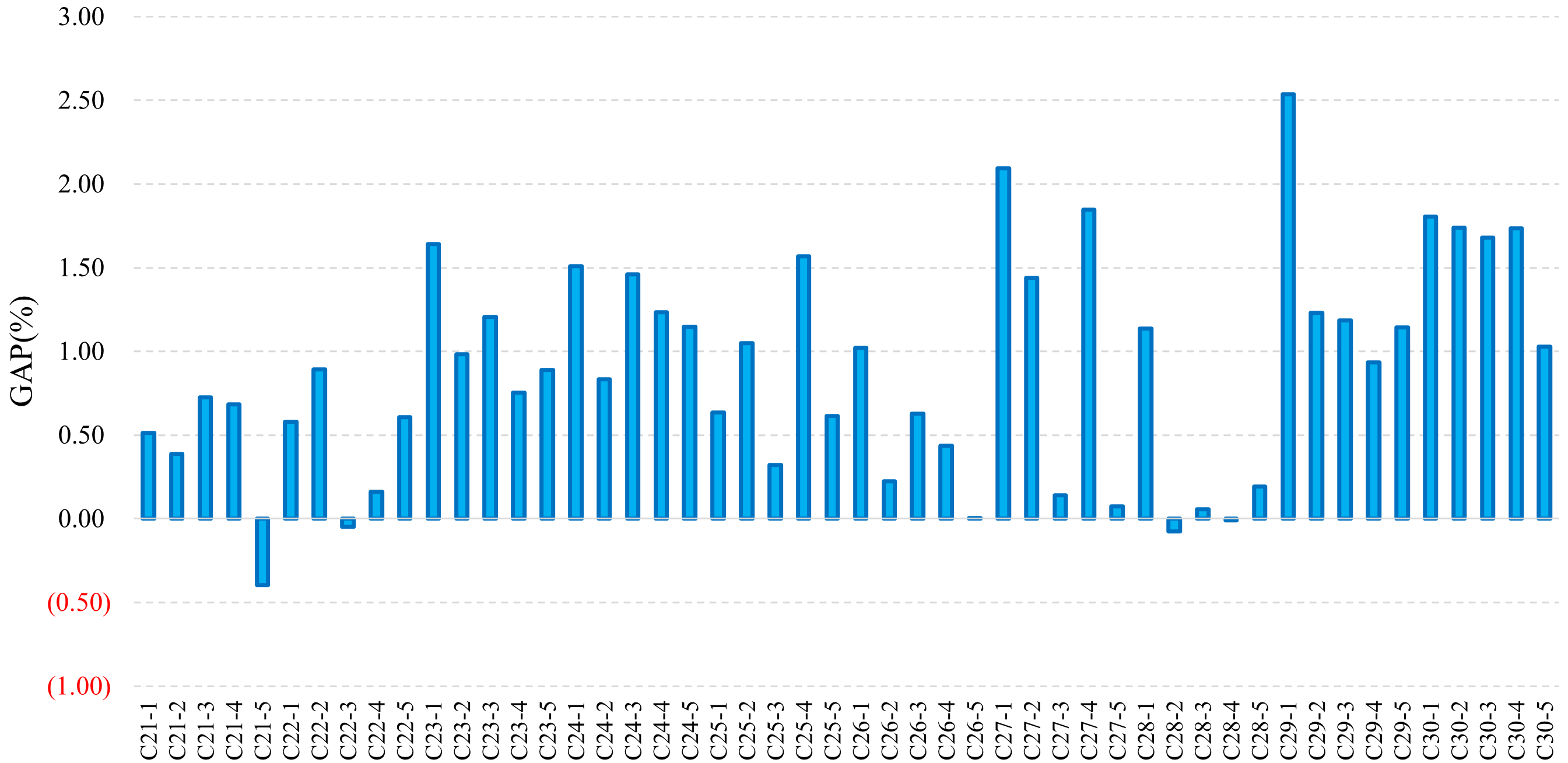}
		\caption{Type \text{II}: The last 50 instances}
	\end{subfigure}
	
	\vspace{3mm}
	\caption{The gap between the results obtained by the AILS-VND-SOCPI and the best-known solutions}
	\label{fig3}
\end{figure}

\begin{figure}[htb]
	\centering
	\begin{subfigure}[t]{0.7\linewidth}
		\captionsetup{justification=centering} 
		\includegraphics[width=\linewidth]{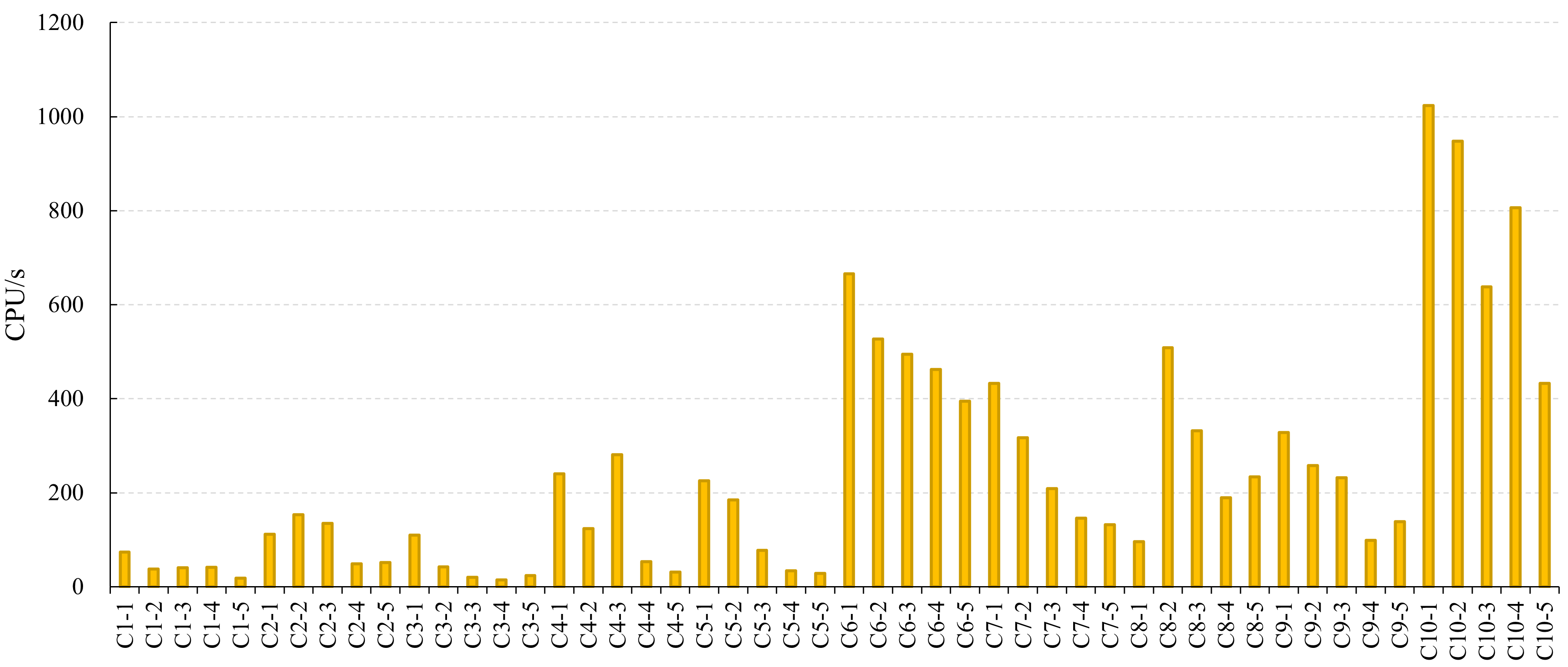}
		\caption{Type \text{I} }
	\end{subfigure}
	\vspace{3mm}

	\begin{subfigure}[t]{0.7\linewidth}
		\captionsetup{justification=centering}
		\includegraphics[width=\linewidth]{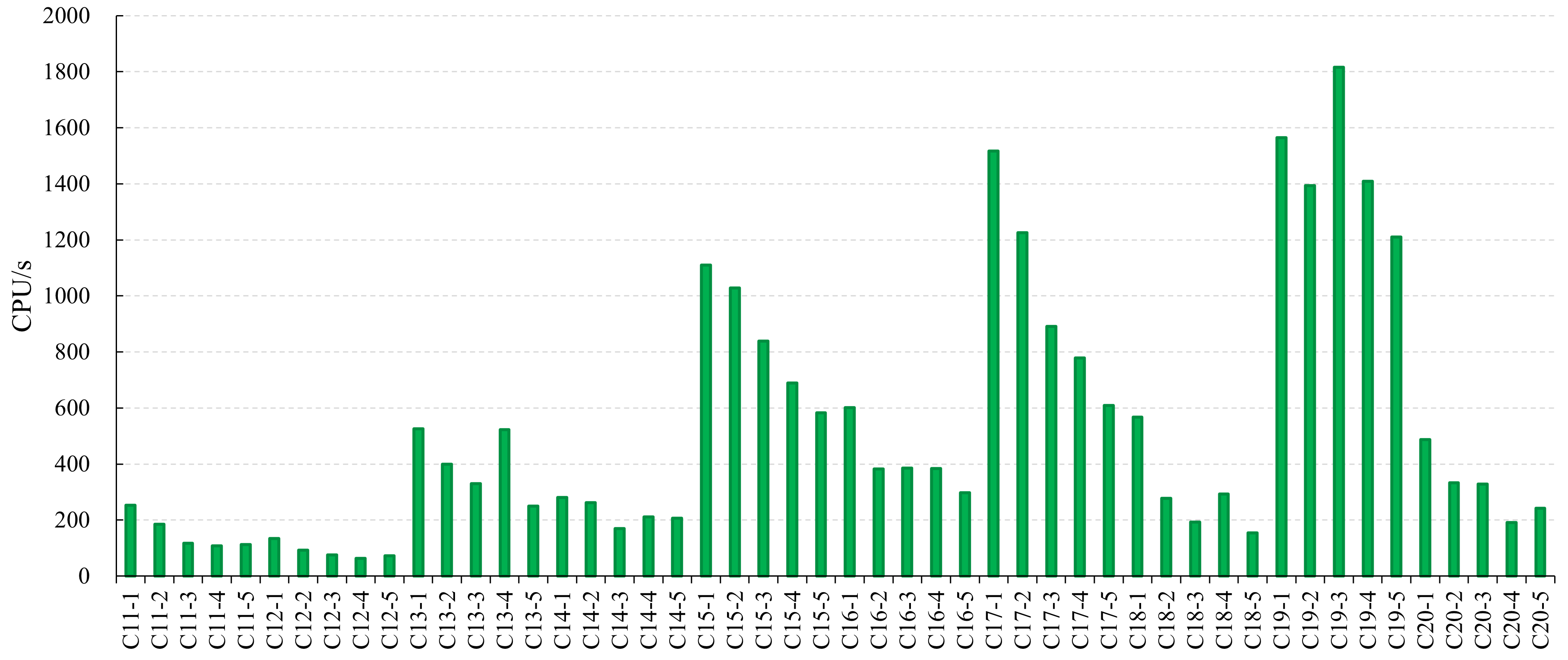}
		\caption{Type \text{II}: The first 50 instances}
	\end{subfigure}
	\vspace{3mm} 
	
	\begin{subfigure}[t]{0.7\linewidth}
		\captionsetup{justification=centering} 
		\includegraphics[width=\linewidth]{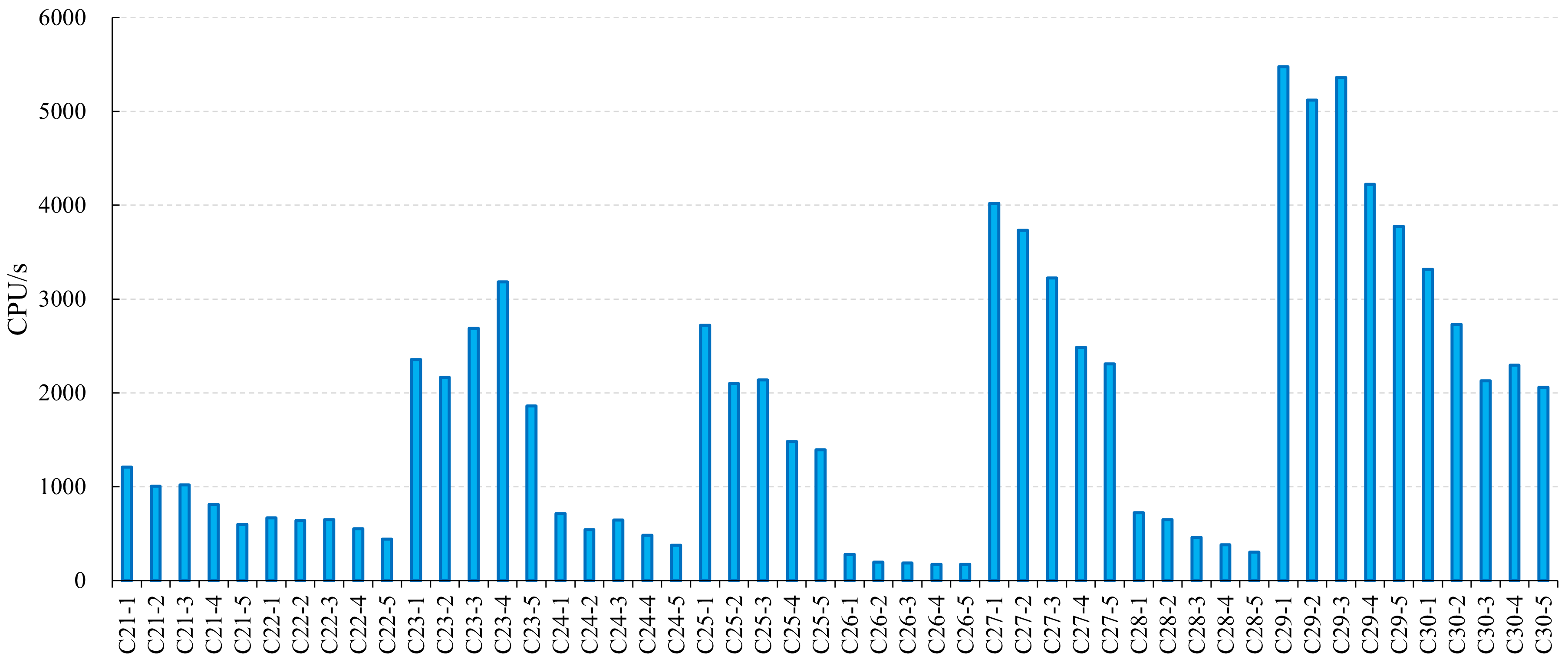}
		\caption{Type \text{II}: The last 50 instances}
	\end{subfigure}
	
	\vspace{3mm}
	\caption{The runtime of the AILS-VND-SOCPI}
	\label{fig4}
\end{figure}

AILS-VND-SOCPI is run 20 times to solve each instance, as shown in Appendix A Tables A.1-A.2. The gap between the results obtained by AILS-VND-SOCPI and the best-known solutions is displayed in Fig.~\ref{fig3}. In addition, the algorithm's average runtime is also provided in Fig.~\ref{fig4}. The gap can be calculated as follows:

\begin{equation}\label{Eq:GAP}
	\begin{array}{cc}
		\text{GAP} = \displaystyle\frac{f(\text{AILS-VND-SOCPI}) - f(\text{B\&C})}{f(\text{B\&C})} \times 100\%.
	\end{array}
\end{equation}

In Fig.~\ref{fig3}(a), the best solutions generated by AILS-VND-SOCPI have a gap within 1\% to the best-known solutions for the 50 instances of Type \text{I}, fully demonstrating the excellent optimization performance of the proposed algorithm. Among these, 41 instances have been verified to reach optimality, with AILS-VND-SOCPI successfully identifying the optimal solution for 24 instances (i.e., GAP = 0\%), accounting for 61\%. Furthermore, for instances C6-3, C6-5, and C7-5, the solutions obtained by AILS-VND-SOCPI outperform the best-known solutions, further demonstrating the algorithm's superior optimization performance. As shown in Fig.~\ref{fig4}(a), AILS-VND-SOCPI can solve all instances of Type \text{I} within 20 minutes. In contrast, B\&C requires 50 to 120 minutes to solve short-range UAV routing problems (i.e., instances Cx-4 to Cx-6) \citep{campbell2023multi}. Although B\&C can produce higher-quality solutions for each instance, it consumes significantly more computational times than AILS-VND-SOCPI. This comparison clearly highlights the superior efficiency of AILS-VND-SOCPI, which substantially reduces computational costs while maintaining solution quality.

Compared to Type \text{I}, Type \text{II} contains more required edges with denser distribution, substantially increasing graph complexity and leading to a dramatic expansion of the search space for the routing problem. As the number of edges increases, the dependencies and connectivity among routes become more intricate. The selection of a single edge can affect the overall solution structure, making it essential to carefully optimize edge routing which is more difficult. In addition, another notable feature of Type \text{II} is that the required edges are generally longer, making it difficult for a single UAV to cover adjacent long edges within the same route. Consequently, how to effectively allocate these long required edges across different routes while minimizing the total route length has emerged as a crucial and urgent challenge.

As shown in Fig.~\ref{fig3}(b) and (c), for the 100 Type \text{II} instances, the best solutions obtained by the AILS-VND-SOCPI exhibit a gap of no more than 2.5\% compared to the best-known solutions. Notably, it outperforms B\&C on 13 instances, demonstrating that the proposed algorithm can better exploit problem-specific features and explore regions of the solution space that B\&C fails to reach.

%

\subsection{Investigation of main mechanisms in AILS-VND-SOCP}
\label{subsec5.2}

\subsubsection{Comparison between with and without disk neighborhoods}
\label{subsubsec5.2.1}

Comparative experiments are conducted on the aforementioned 150 instances to assess the effectiveness of disk neighborhoods. The radius of each disk neighborhood is 50. AILS-VND-SOCP and AILS-VND-SOCP without disk neighborhood (AILS-VND-SOCPI) are run 20 times to solve each instance, as shown in Appendix A Tables A.3-A.4.

\begin{figure}[!hb]
	\centering
	\begin{subfigure}[t]{0.7\linewidth}
		\captionsetup{justification=centering} 
		\includegraphics[width=\linewidth]{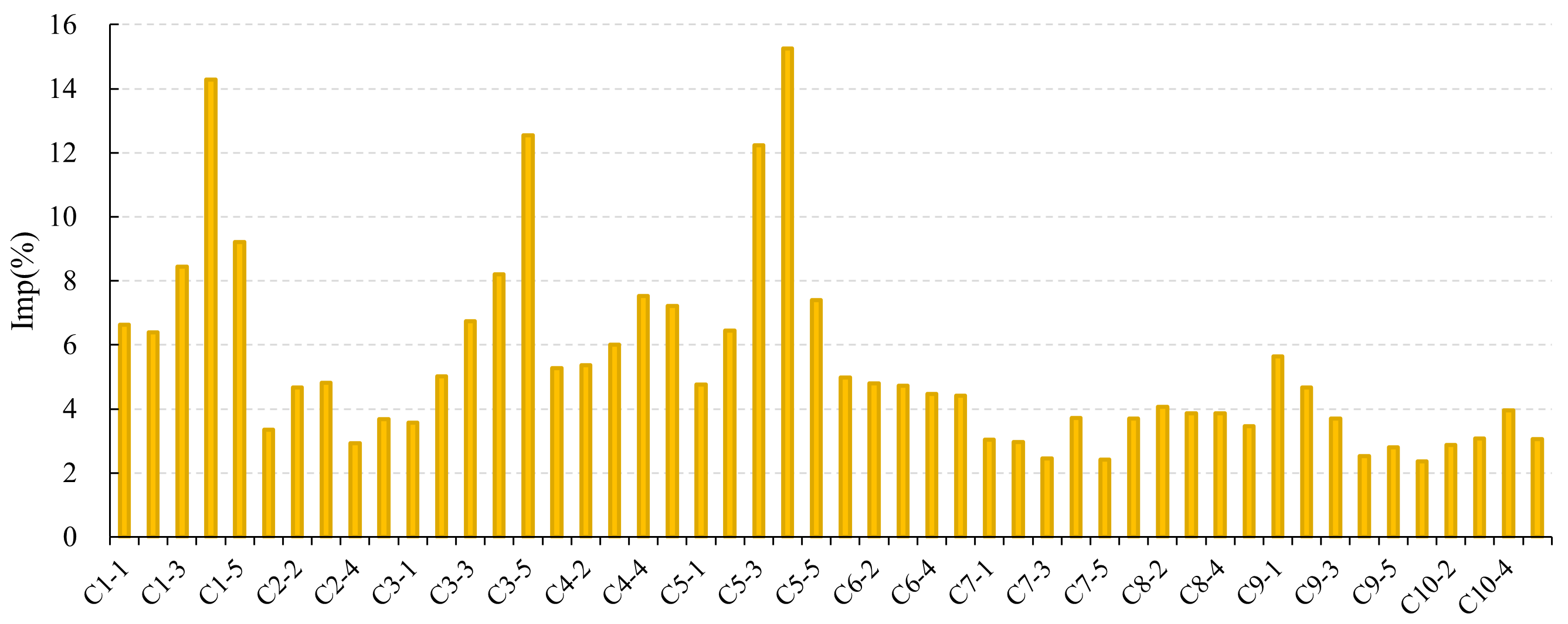}
		\caption{Type \text{I} }
	\end{subfigure}
	\vspace{3mm} 
	
	\begin{subfigure}[t]{0.7\linewidth}
		\captionsetup{justification=centering} 
		\includegraphics[width=\linewidth]{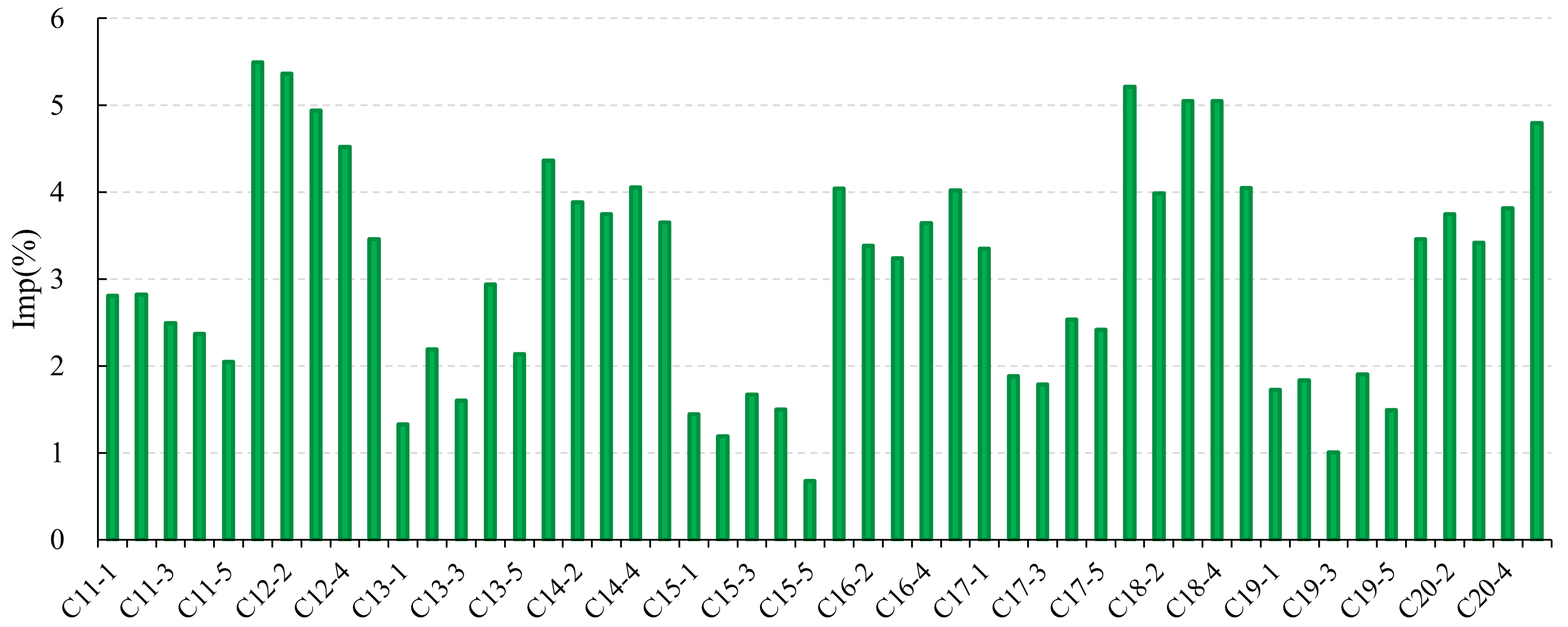}
		\caption{Type \text{II}: The first 50 instances}
	\end{subfigure}
	\vspace{3mm} 
	
	\begin{subfigure}[t]{0.7\linewidth}
		\captionsetup{justification=centering} 
		\includegraphics[width=\linewidth]{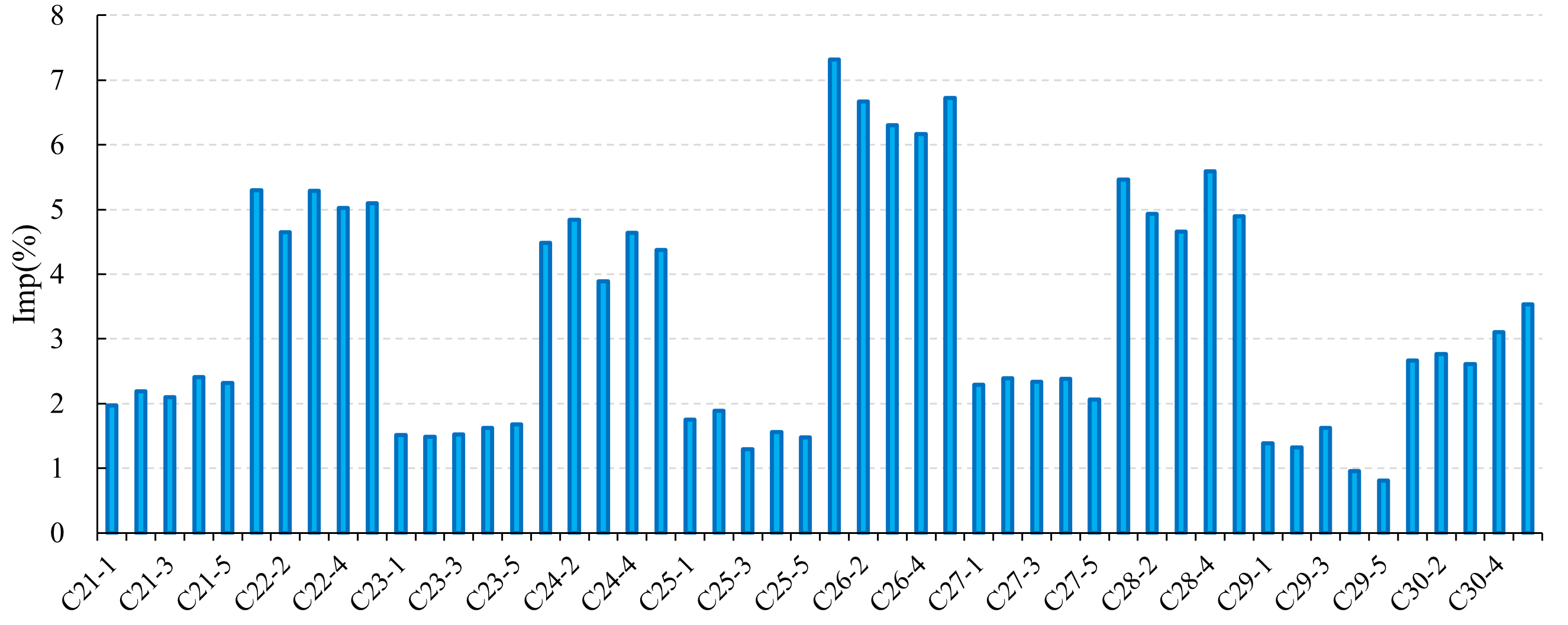}
		\caption{Type \text{II}: The last 50 instances}
	\end{subfigure}
	
	\vspace{3mm}
	\caption{The saving rate of flight distance relative to the case without disk neighborhoods}
	\label{fig6}
\end{figure}

To clearly illustrate the impact of the disk neighborhoods on the final flight distance, the saving rate of the flight distance is presented in Fig.~\ref{fig6}. The saving rate represents the deviation between the average distances produced by AILS-VND-SOCPI and AILS-VND-SOCP, calculated as follows:

\begin{equation}\label{Eq:Imp}
	\begin{array}{cc}
		Imp = \displaystyle\frac{f(\textit{AILS-VND-SOCPI}) - f(\textit{AILS-VND-SOCP})}{f(\textit{AILS-VND-SOCPI})} \times 100\%.
	\end{array}
\end{equation}

\begin{figure}[!hb]
	\centering
	\begin{subfigure}[t]{0.7\linewidth}
		\captionsetup{justification=centering} 
		\includegraphics[width=\linewidth]{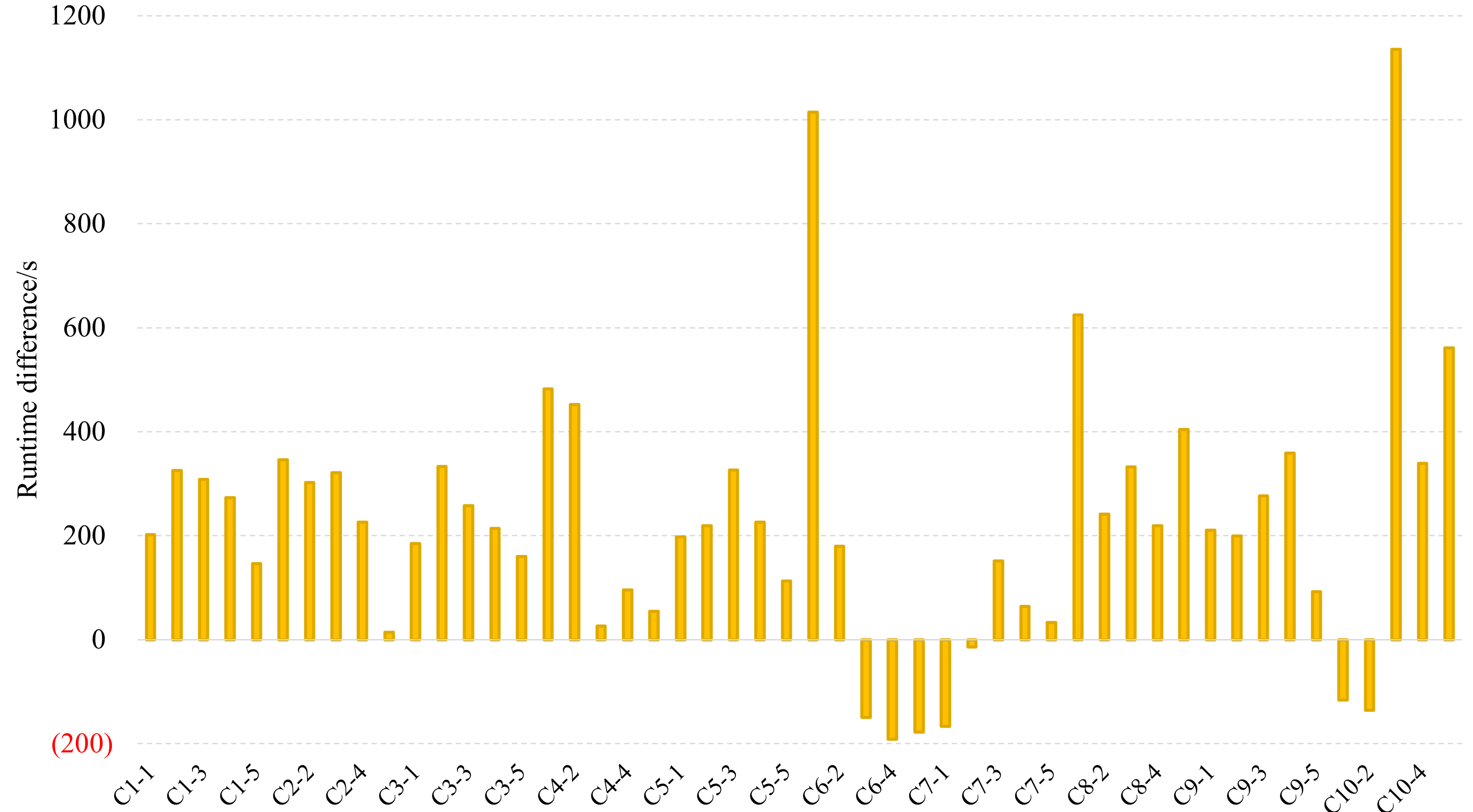}
		\caption{Type \text{I} }
	\end{subfigure}
	\vspace{3mm} 
	
	\begin{subfigure}[t]{0.7\linewidth}
		\captionsetup{justification=centering} 
		\includegraphics[width=\linewidth]{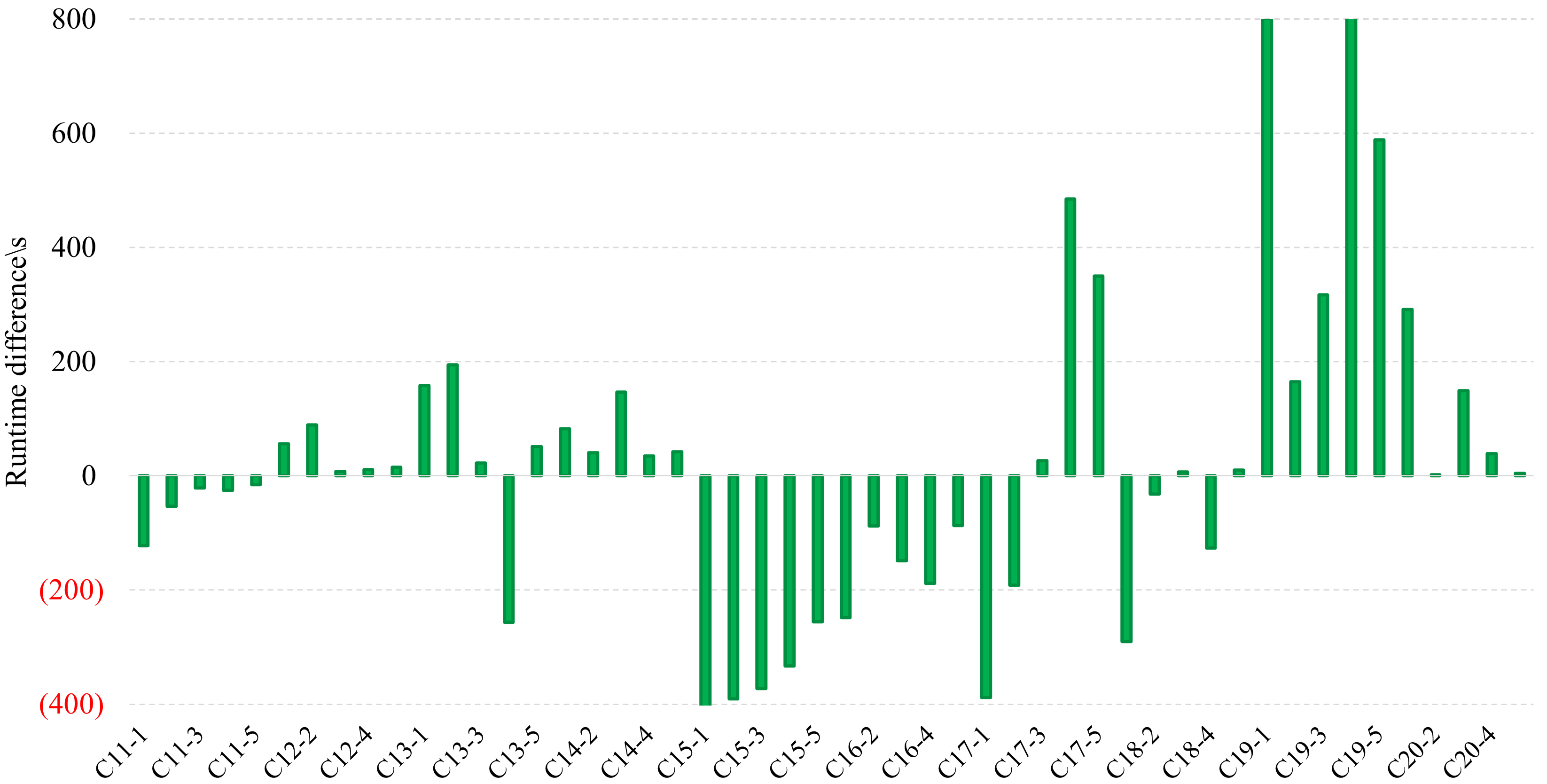}
		\caption{Type \text{II}: The first 50 instances}
	\end{subfigure}
	\vspace{3mm} 
	
	\begin{subfigure}[t]{0.7\linewidth}
		\captionsetup{justification=centering} 
		\includegraphics[width=\linewidth]{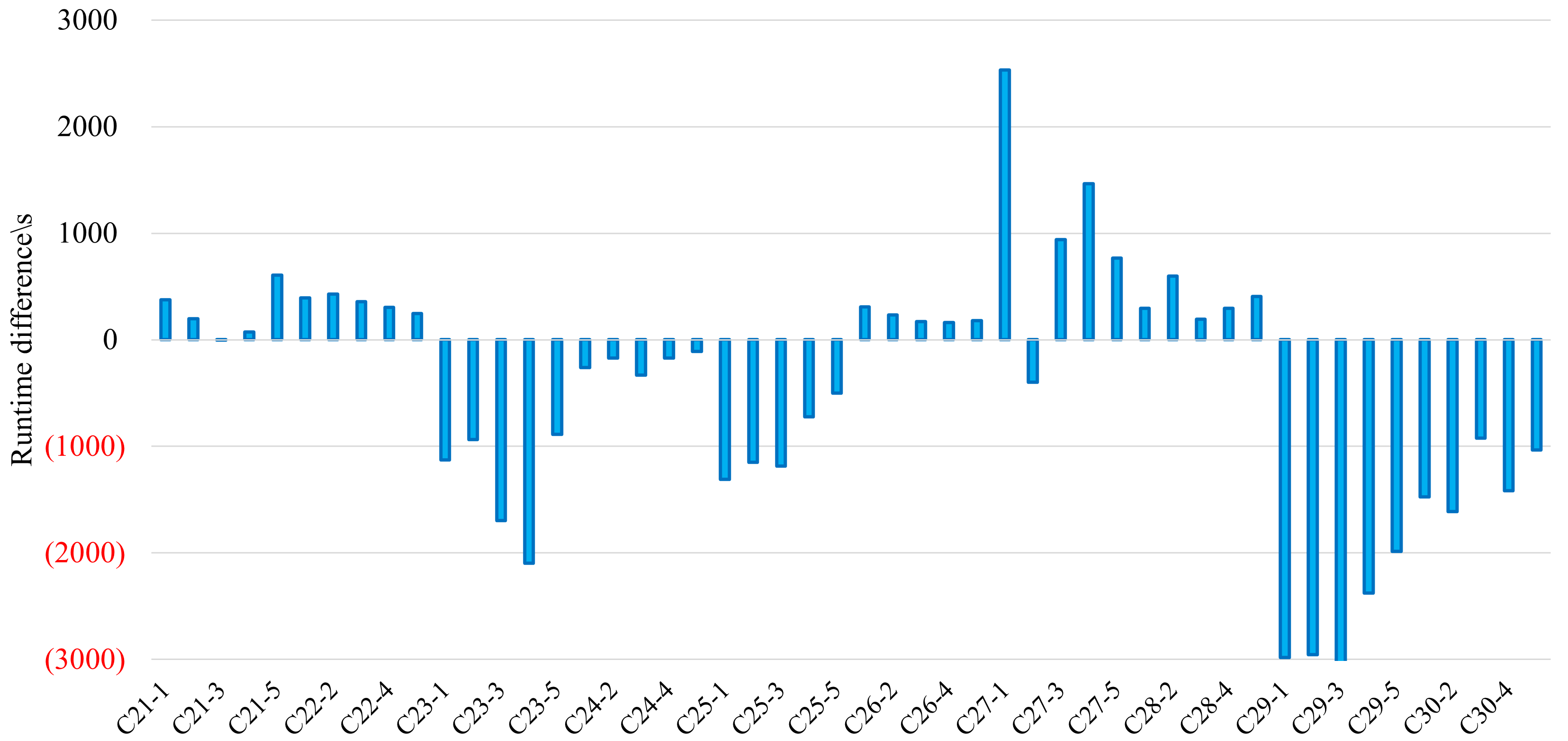}
		\caption{Type \text{II}: The last 50 instances}
	\end{subfigure}
	
	\vspace{3mm}
	\caption{The runtime of AILS-VND-SOCP}
	\label{fig7}
\end{figure}

\begin{landscape}
	\begin{figure}[]
		\centering
		\begin{tabular}{@{}c@{}c@{}c@{}}
			\begin{subfigure}[t]{0.33\linewidth}
				\centering
				\captionsetup{justification=centering} 
				\includegraphics[width=0.9\linewidth]{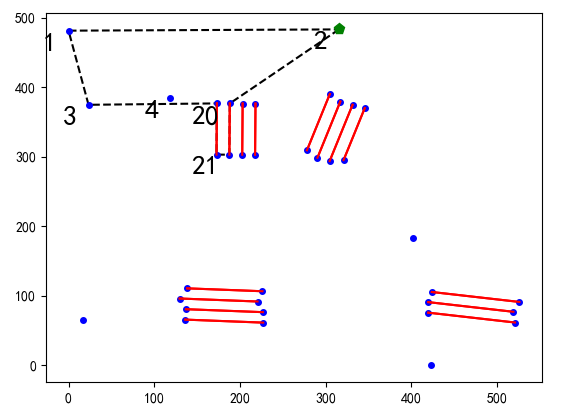}
				\caption{Without disk neighborhoods: UAV1}
			\end{subfigure}
			
			\begin{subfigure}[t]{0.33\linewidth}
				\centering
				\captionsetup{justification=centering} 
				\includegraphics[width=0.9\linewidth]{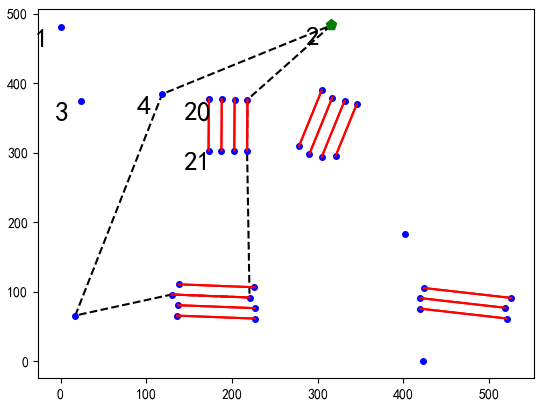}
				\caption{Without disk neighborhoods: UAV2}
			\end{subfigure}
			
			\begin{subfigure}[t]{0.33\linewidth}
				\centering
				\captionsetup{justification=centering} 
				\includegraphics[width=0.9\linewidth]{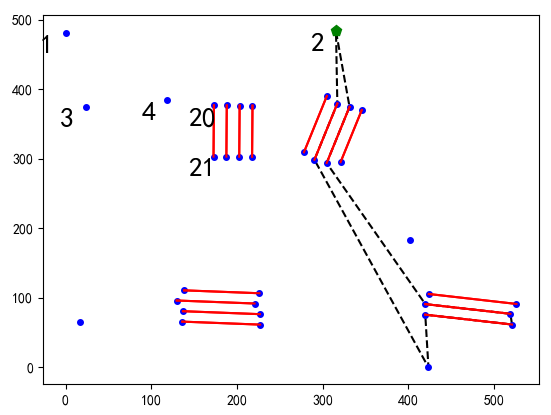}
				\caption{Without disk neighborhoods: UAV3}
			\end{subfigure}\\
			
			\begin{subfigure}[t]{0.33\linewidth}
				\centering
				\captionsetup{justification=centering} 
				\includegraphics[width=0.9\linewidth]{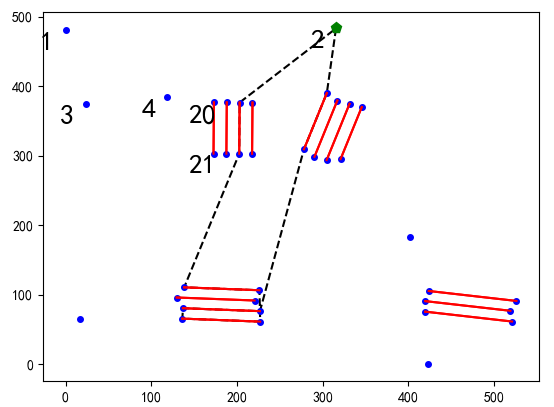}
				\caption{Without disk neighborhoods: UAV4}
			\end{subfigure}
			
			\begin{subfigure}[t]{0.33\linewidth}
				\centering
				\captionsetup{justification=centering} 
				\includegraphics[width=0.9\linewidth]{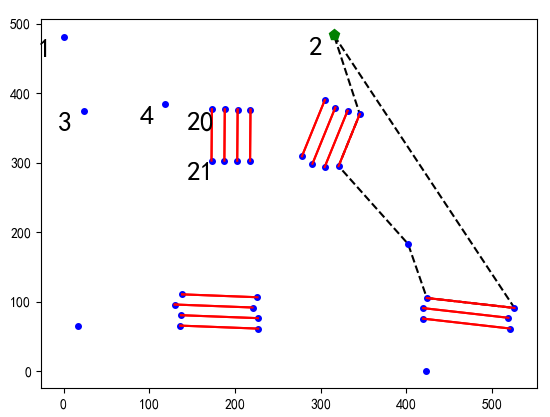}
				\caption{Without disk neighborhoods: UAV5}
			\end{subfigure}
			
			\begin{subfigure}[t]{0.33\linewidth}
				\centering
				\captionsetup{justification=centering} 
				\includegraphics[width=0.9\linewidth]{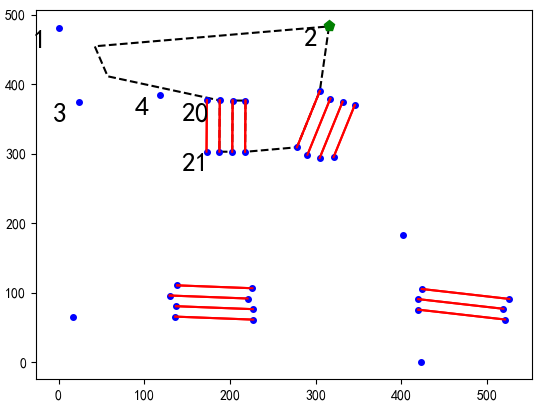}
				\caption{With disk neighborhoods: UAV1}
			\end{subfigure}\\
			
			\begin{subfigure}[t]{0.33\linewidth}
				\centering
				\captionsetup{justification=centering} 
				\includegraphics[width=0.9\linewidth]{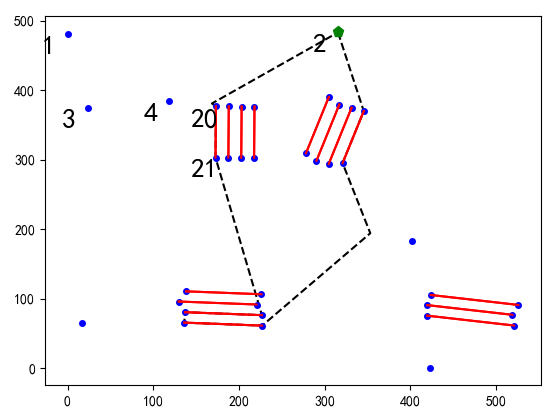}
				\caption{With disk neighborhoods: UAV2}
			\end{subfigure}
			
			\begin{subfigure}[t]{0.33\linewidth}
				\centering
				\captionsetup{justification=centering} 
				\includegraphics[width=0.9\linewidth]{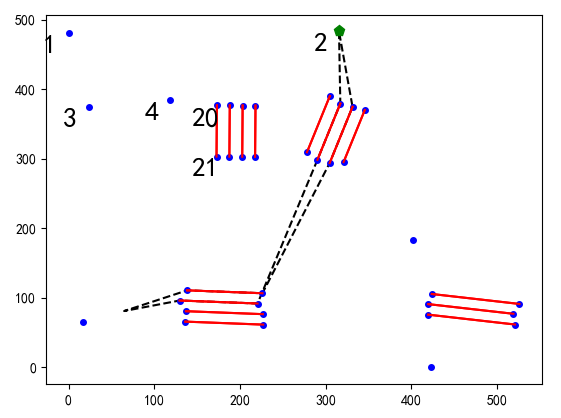}
				\caption{With disk neighborhoods: UAV3}
			\end{subfigure}
			
			\begin{subfigure}[t]{0.33\linewidth}
				\centering
				\captionsetup{justification=centering} 
				\includegraphics[width=0.9\linewidth]{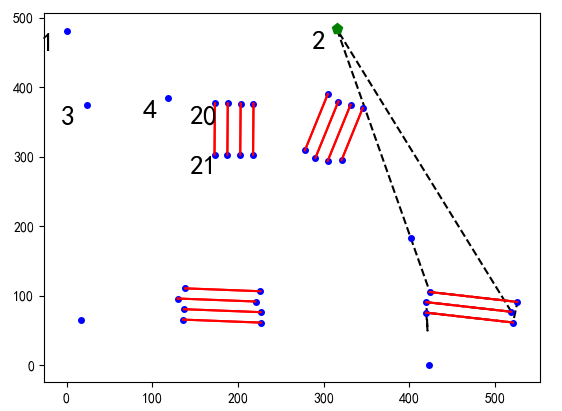}
				\caption{With disk neighborhoods: UAV4}
			\end{subfigure}\\
		\end{tabular}
		\vspace{2mm}
		\caption{The solutions of C1-4 with and without disk neighborhoods}
		\label{fig8} 
	\end{figure}
\end{landscape}

\begin{figure}[!h]
	\centering
	\begin{subfigure}[t]{0.6\linewidth}
		\captionsetup{justification=centering} 
		\includegraphics[width=\linewidth]{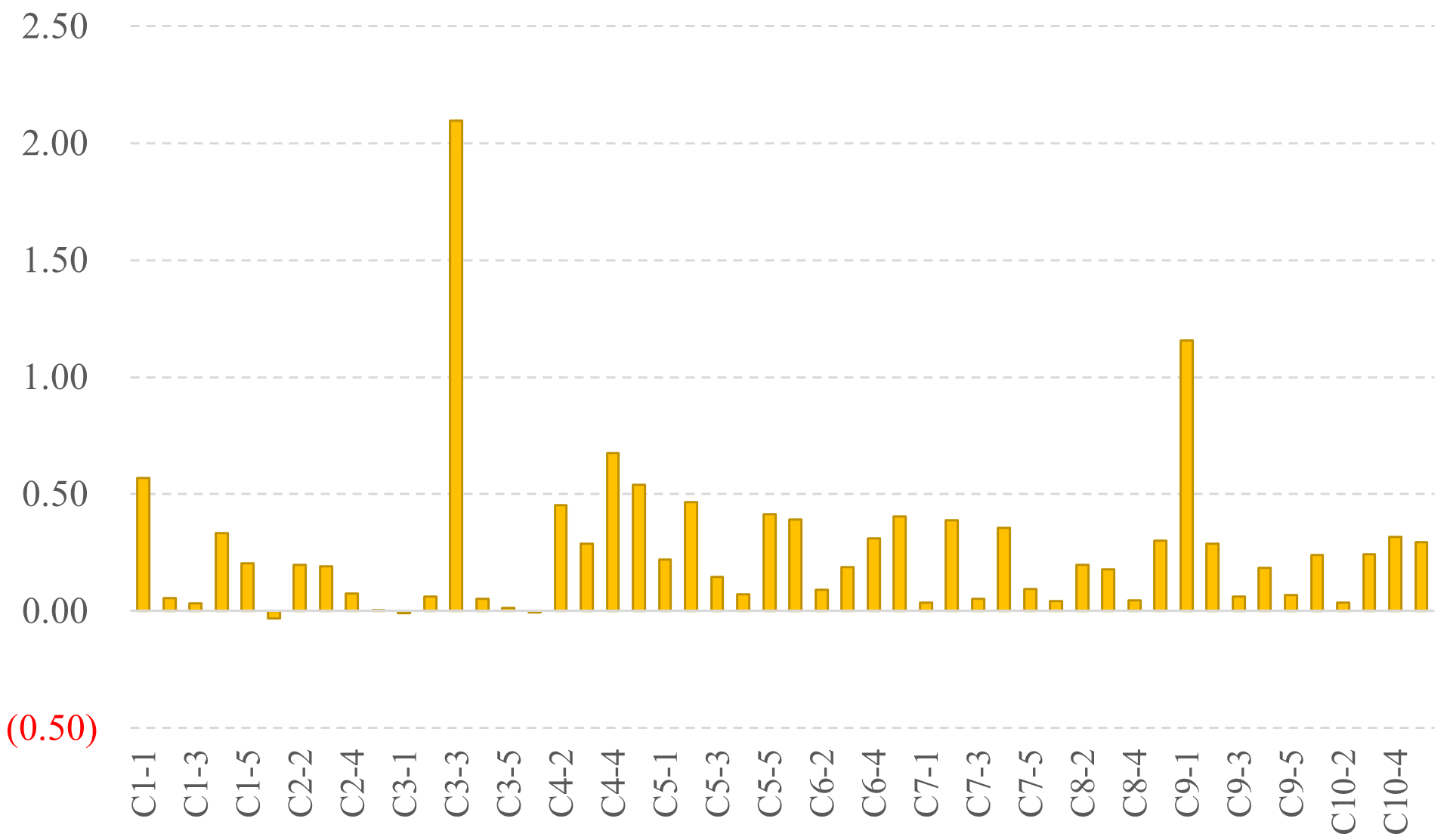}
		\caption{Type \text{I} }
	\end{subfigure}
	\vspace{3mm}

	\begin{subfigure}[t]{0.6\linewidth}
		\captionsetup{justification=centering} 
		\includegraphics[width=\linewidth]{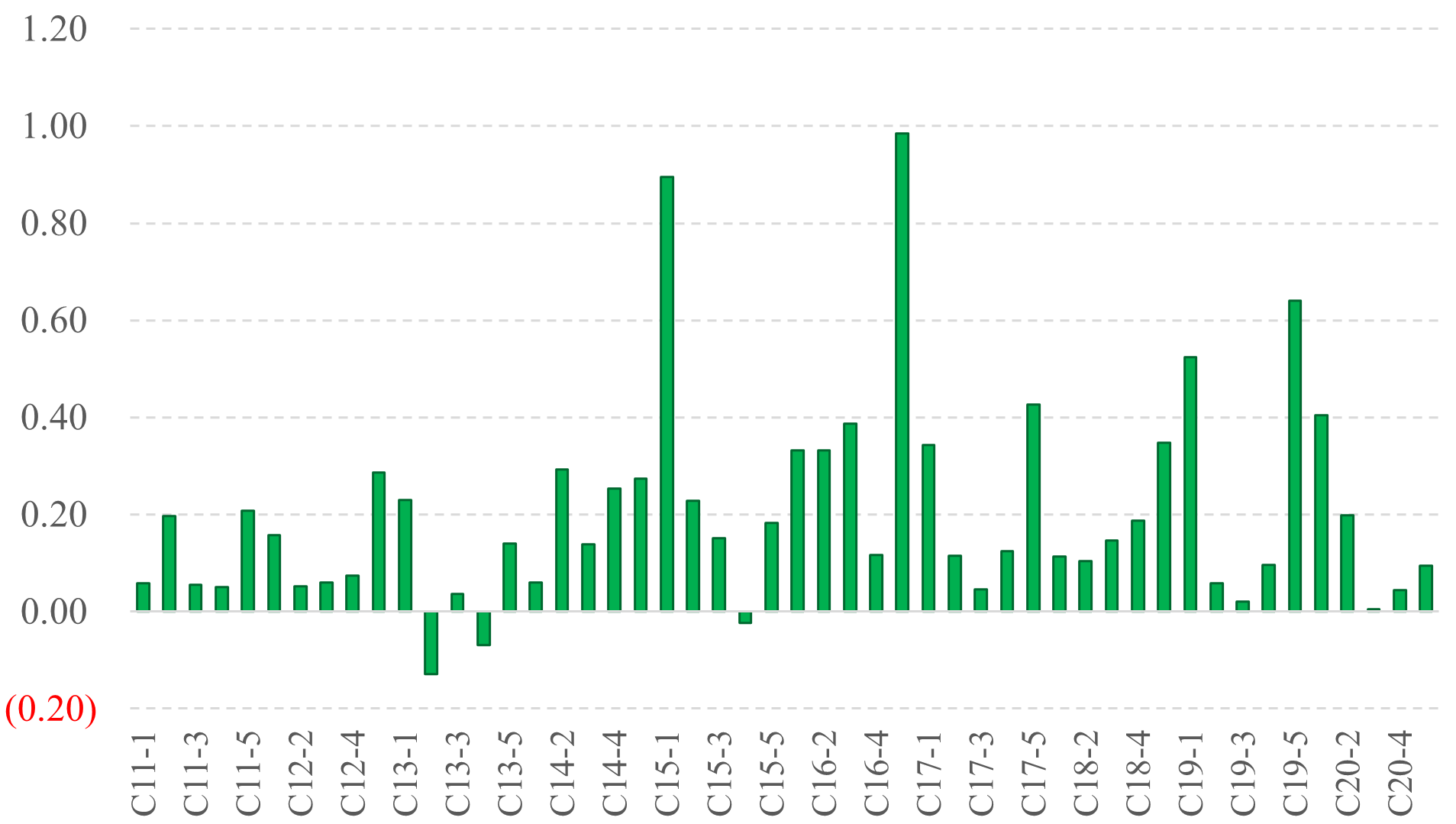}
		\caption{Type \text{II}: The first 50 instances}
	\end{subfigure}
	\vspace{3mm}
	
	\begin{subfigure}[t]{0.6\linewidth}
		\captionsetup{justification=centering} 
		\includegraphics[width=\linewidth]{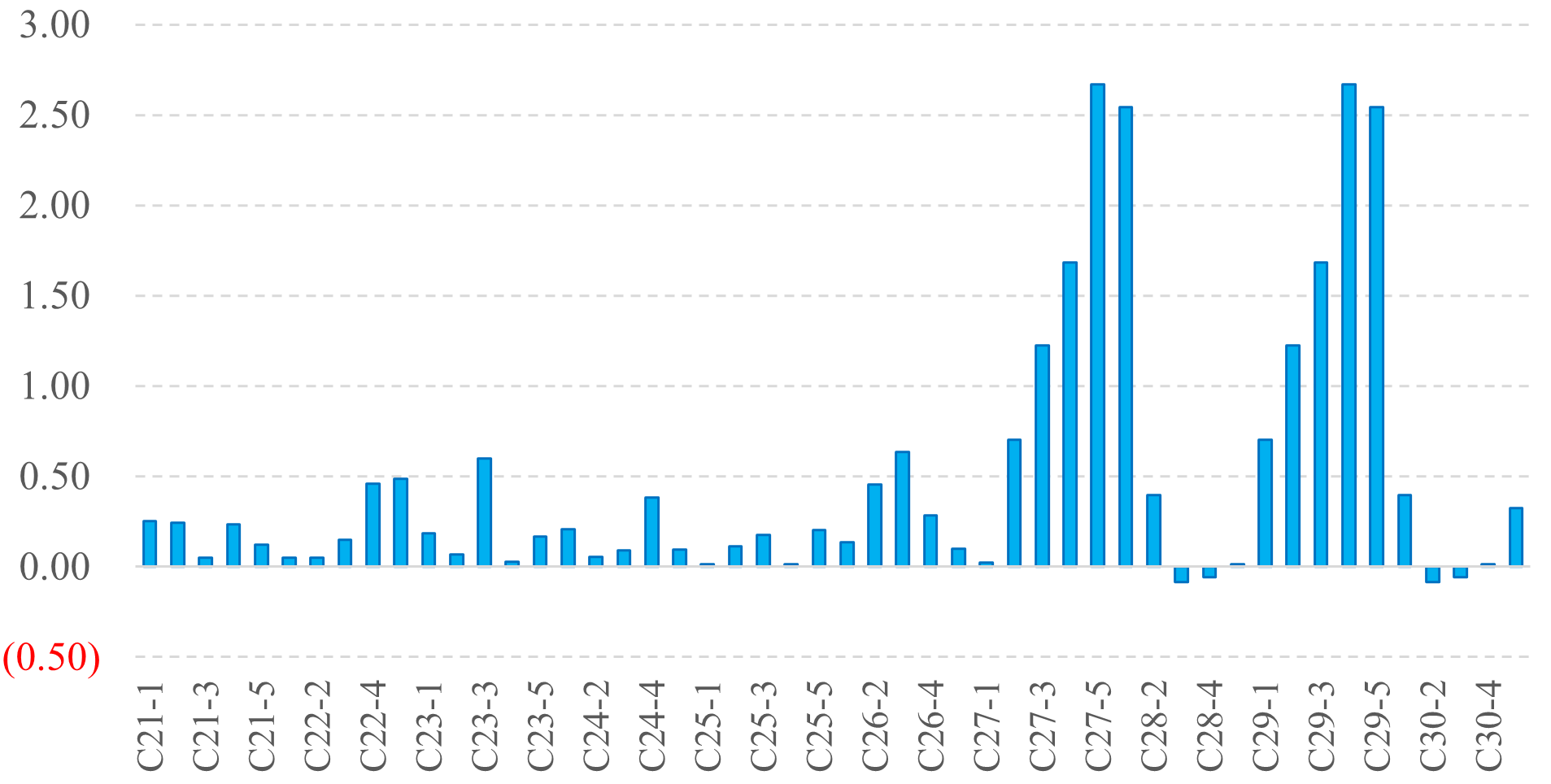}
		\caption{Type \text{II}: The last 50 instances}
	\end{subfigure}
	
	\vspace{3mm}
	\caption{The saving rate of flight distance relative to the case without re-increasing acceptance threshold}
	\label{fig9}
\end{figure}

As shown in Fig.~\ref{fig6}, among the 150 instances, the coverage monitoring strategy (considering disk neighborhoods) can effectively reduce the total flight distance of UAVs, especially for Type \text{I} instances. For example, the saving rate of the flight distance is 14.28\% for instance C1-4. This can be attributed to the spatial distribution differences between Type~\text{I} and Type~\text{II} instances. In Type~\text{I} instances, the required nodes are more sparsely distributed, and the required edges are fewer and shorter. Consequently, the flight distance between required nodes accounts for a larger proportion of the total flight distance. Thus, the coverage monitoring strategy can significantly reduce the flight distance between required nodes, thereby decreasing the overall flight distance. Furthermore, in some instances, this strategy can also reduce the number of used UAVs. For example, as shown in Tables S3-S4, the number of used UAVs is decreased in instances C1-4, C2-4, C2-5, C3-4, C3-5, and C5-5. These results demonstrate that the coverage monitoring strategy can not only lower the flight distance but also effectively reduce the number of used UAVs.

As shown in Fig.~\ref{fig7}, the coverage monitoring strategy can increase the runtime for most instances, primarily due to the additional computational resources required for representative point optimization. However, in some instances, this strategy can change the topology of the optimal solution, thus facilitating the convergence of the algorithm and ultimately producing better solutions within a shorter runtime.

To better demonstrate the benefits of the coverage monitoring strategy, a comparison between the results with and without the strategy for instance C1-4 is presented in Fig.~\ref{fig8}. As shown in Fig.~\ref{fig8}(f), after completing node \#3, the UAV directly flies toward edge target (20, 21). During this process, its route passes over node \#4, which is within the airborne sensor coverage range. As a result, the UAV can cover node \#4 without any additional flight distance. This strategy not only enhances the operational efficiency of UAVs (allowing them to complete more tasks), but also, in some cases, significantly reduces the total number of used UAVs (e.g., from 5 to 4 for instance C1-4).

\subsubsection{Comparison between with and without re-increasing the acceptance threshold}
\label{subsubsec5.2.2}

To verify the efficiency of re-increasing the acceptance threshold, the ALNS-VND-SOCP solves the aforementioned 150 instances, excluding the threshold re-enlargement operator to quantify its impact on performance. The degradation in solution quality---measured by the saving rate (Eq. \ref{Eq:Imp})---serves as an indicator of the operator's importance in optimizing solutions. For comparative analysis, both the standard AILS-VND-SOCP and its modified variant without threshold re-enlargement (AILS-VND-SOCPII) are run 20 times to solve each instance (see Appendix A Tables A.5-A.6). The saving rate of the flight distance is further illustrated in Fig.~\ref{fig9}.

It can be observed that in most instances, re-increasing the acceptance threshold effectively reduces the flight distance, especially for the last 50 instances of Type {II}. This is because these instances contain more required edges with higher density, making the algorithm more prone to being trapped in local optima. By re-increasing the acceptance threshold, AILS-VND-SOCP is able to accept more inferior solutions, thereby enhancing its ability to escape from local optima and explore better regions of the solution space.

\subsection{Sensitivity to the radius of disk neighborhoods}
\label{subsec5.3}

\begin{figure}[htb]
	
	\centering
	\begin{tabular}{@{}c@{}c@{}}
		\begin{subfigure}[t]{0.4\linewidth}
			\captionsetup{justification=centering} 
			\begin{minipage}[b]{1\linewidth}
				\includegraphics[width=1\linewidth]{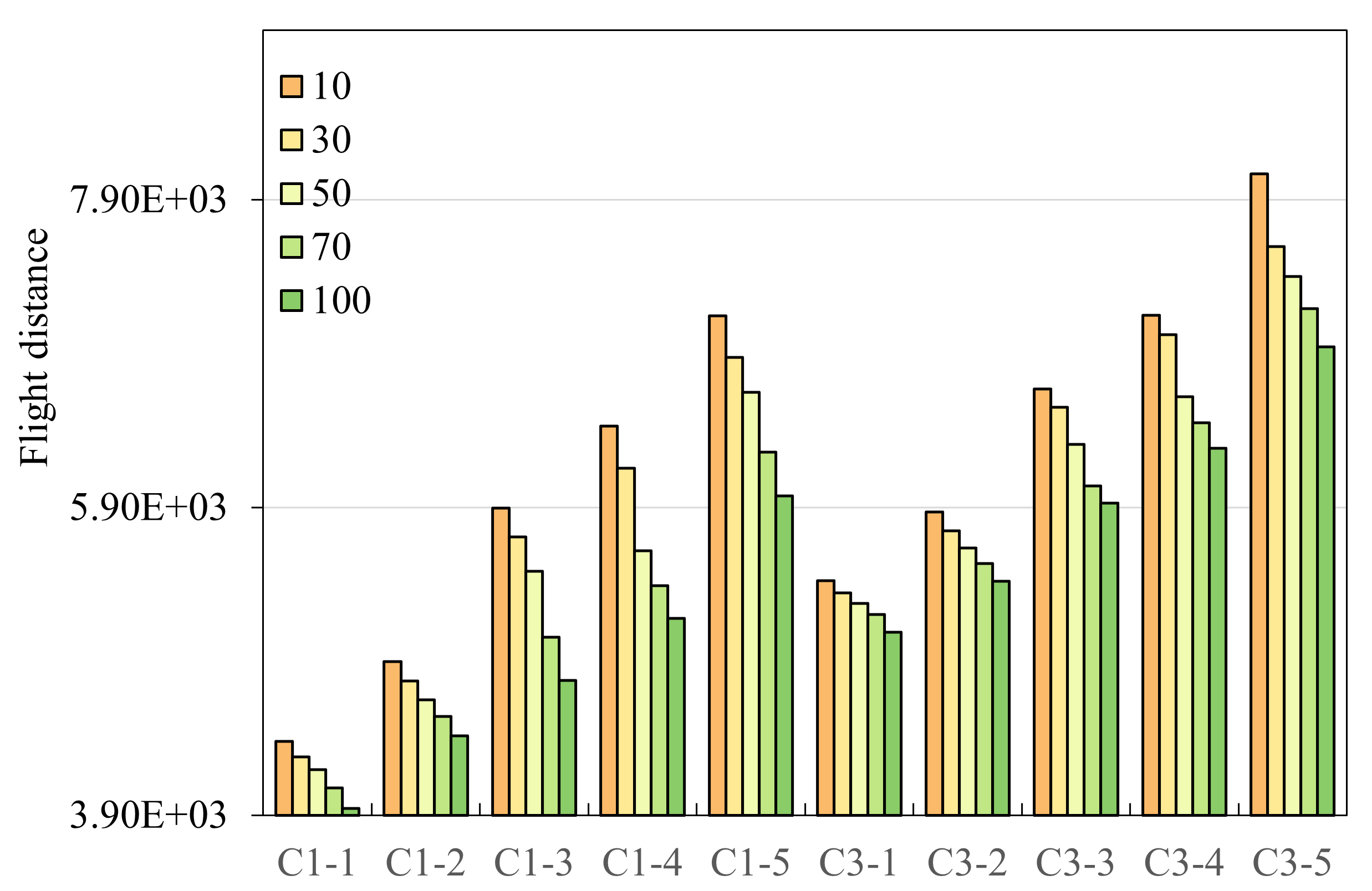}
				\caption{}
			\end{minipage}
		\end{subfigure}
		
		\begin{subfigure}[t]{0.4\linewidth}
			\captionsetup{justification=centering} 
			\begin{minipage}[b]{1\linewidth}
				\includegraphics[width=1\linewidth]{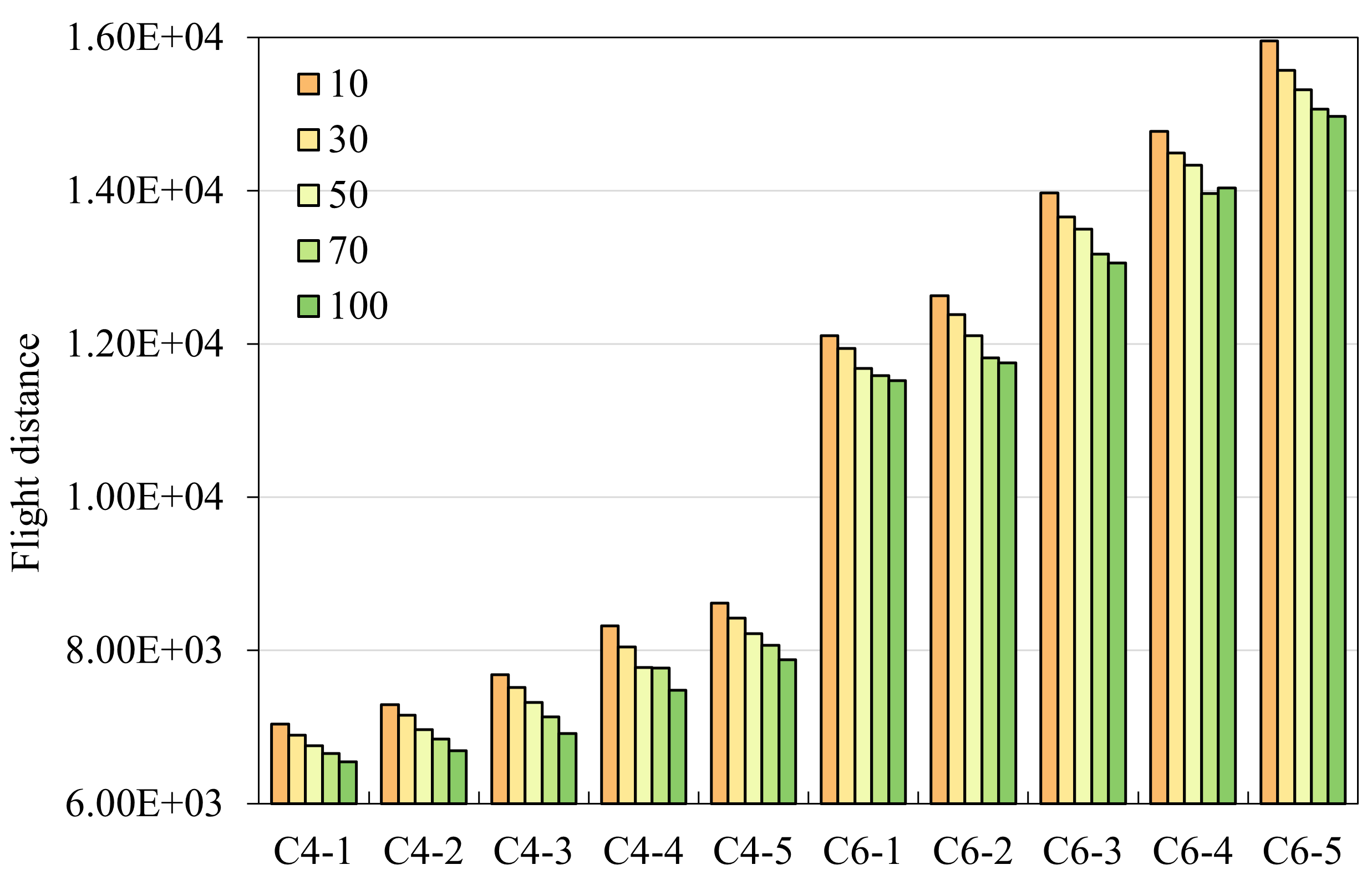}
				\caption{}
			\end{minipage}
		\end{subfigure}\\
		
		\begin{subfigure}[t]{0.4\linewidth}
			\captionsetup{justification=centering} 
			\begin{minipage}[b]{1\linewidth}
				\includegraphics[width=1\linewidth]{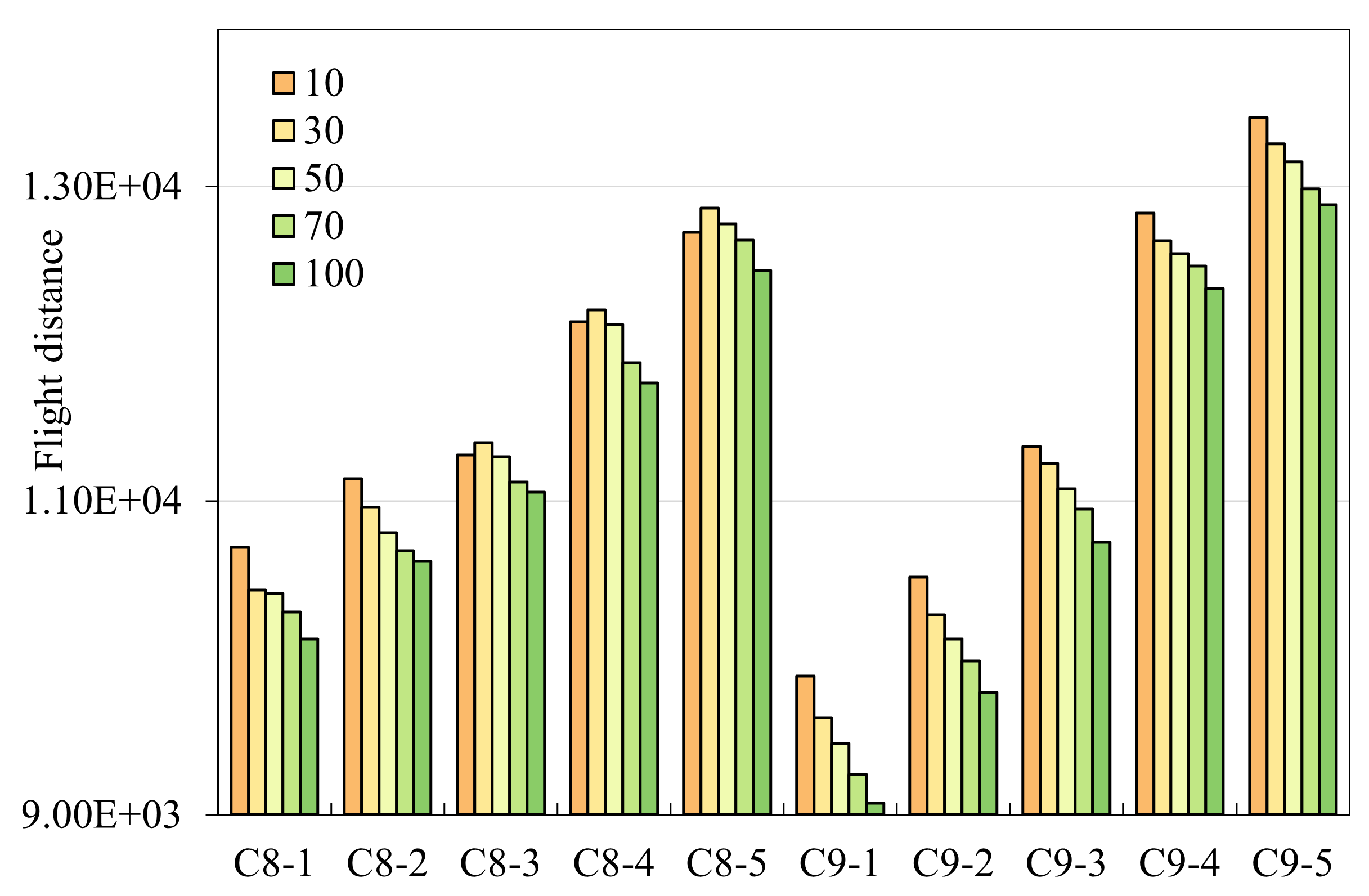}
				\caption{}
			\end{minipage}
		\end{subfigure}
		
		\begin{subfigure}[t]{0.4\linewidth}
			\captionsetup{justification=centering} 
			\begin{minipage}[b]{1\linewidth}
				\includegraphics[width=1\linewidth]{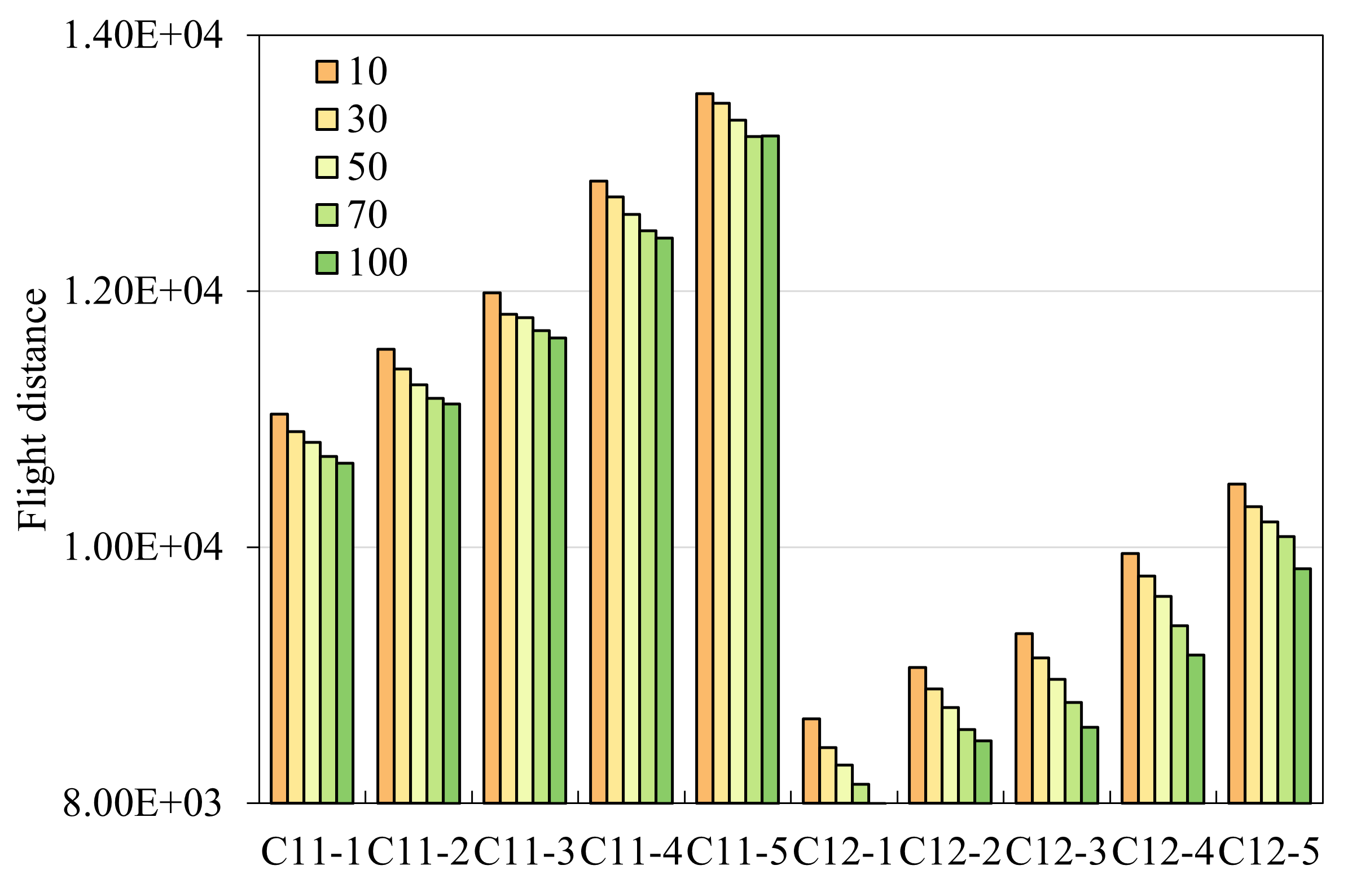}
				\caption{}
			\end{minipage}
		\end{subfigure}\\
		
		\begin{subfigure}[t]{0.4\linewidth}
			\captionsetup{justification=centering} 
			\begin{minipage}[b]{1\linewidth}
				\includegraphics[width=1\linewidth]{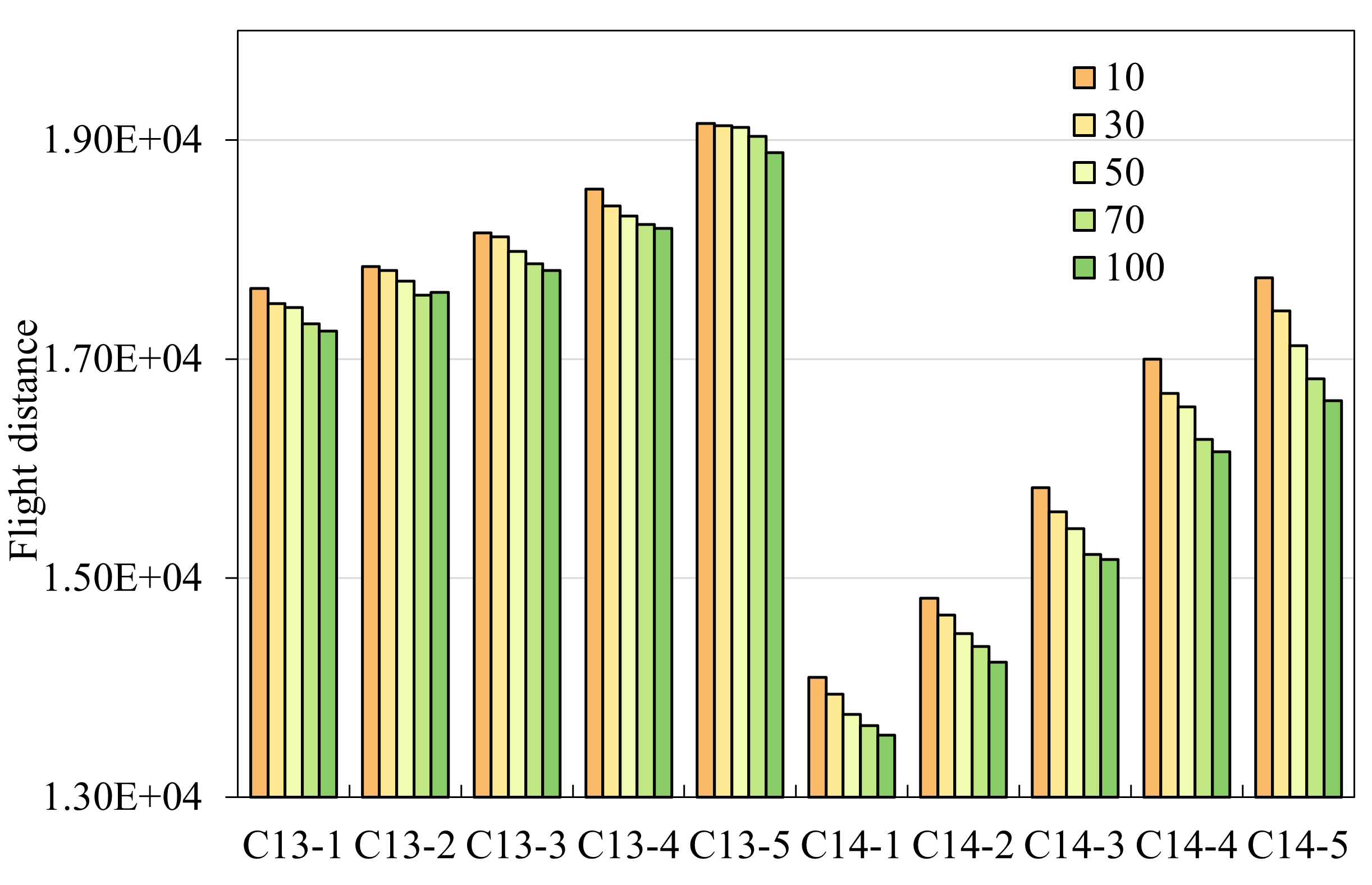}
				\caption{}
			\end{minipage}
		\end{subfigure}
		
		\begin{subfigure}[t]{0.4\linewidth}
			\captionsetup{justification=centering} 
			\begin{minipage}[b]{1\linewidth}
				\includegraphics[width=1\linewidth]{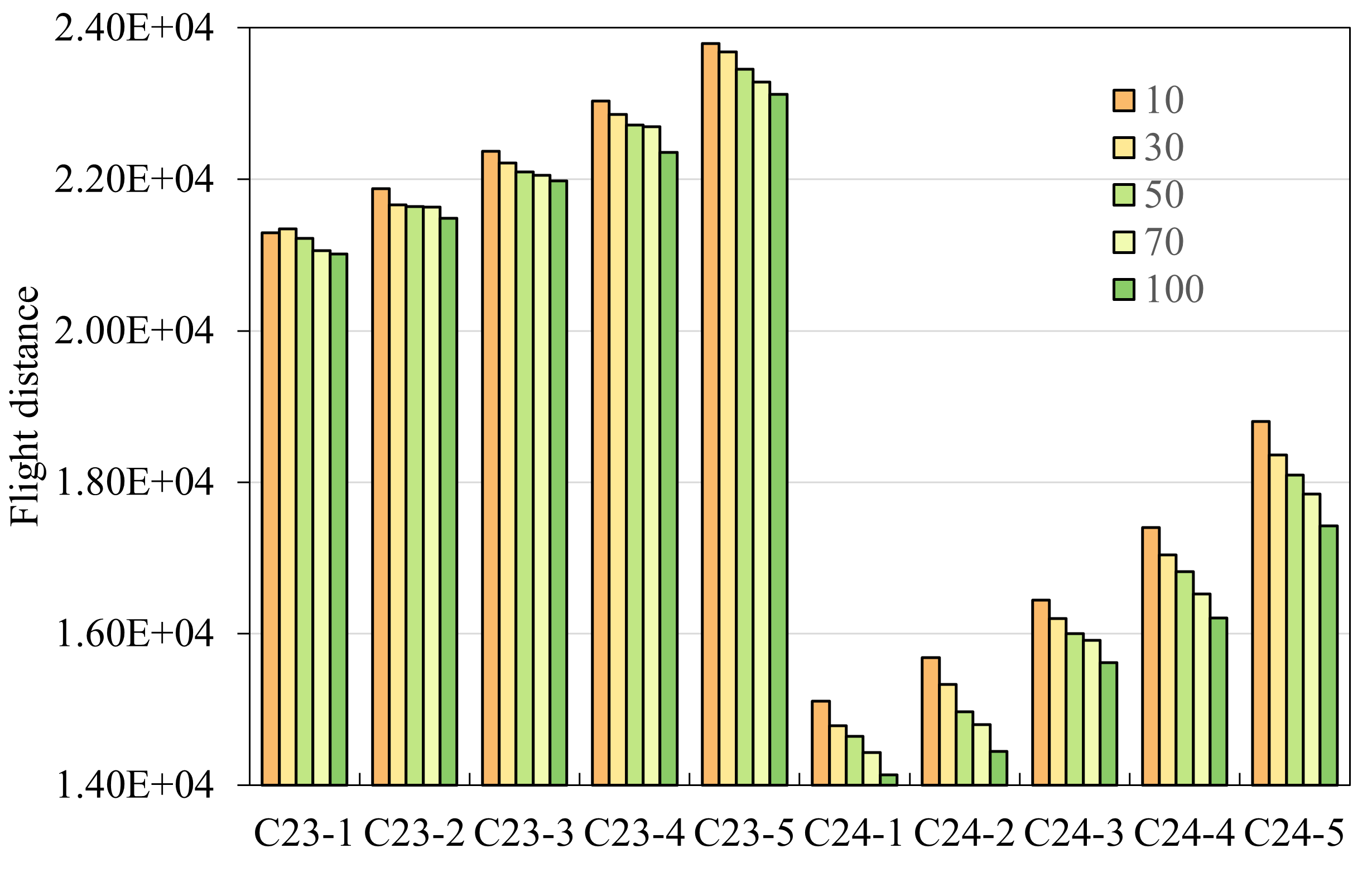}
				\caption{}
			\end{minipage}
		\end{subfigure}\\

	\end{tabular}
	\vspace{-2mm} 
	
	\caption{The flight distance with different radius of disk neighborhoods}
	\label{fig10} 
	\vspace{-3mm}
\end{figure}

\begin{figure}[!h]
	
	\centering
	\begin{tabular}{@{}c@{}c@{}}
		\begin{subfigure}[t]{0.4\linewidth}
			\captionsetup{justification=centering} 
			\begin{minipage}[b]{1\linewidth}
				\includegraphics[width=1\linewidth]{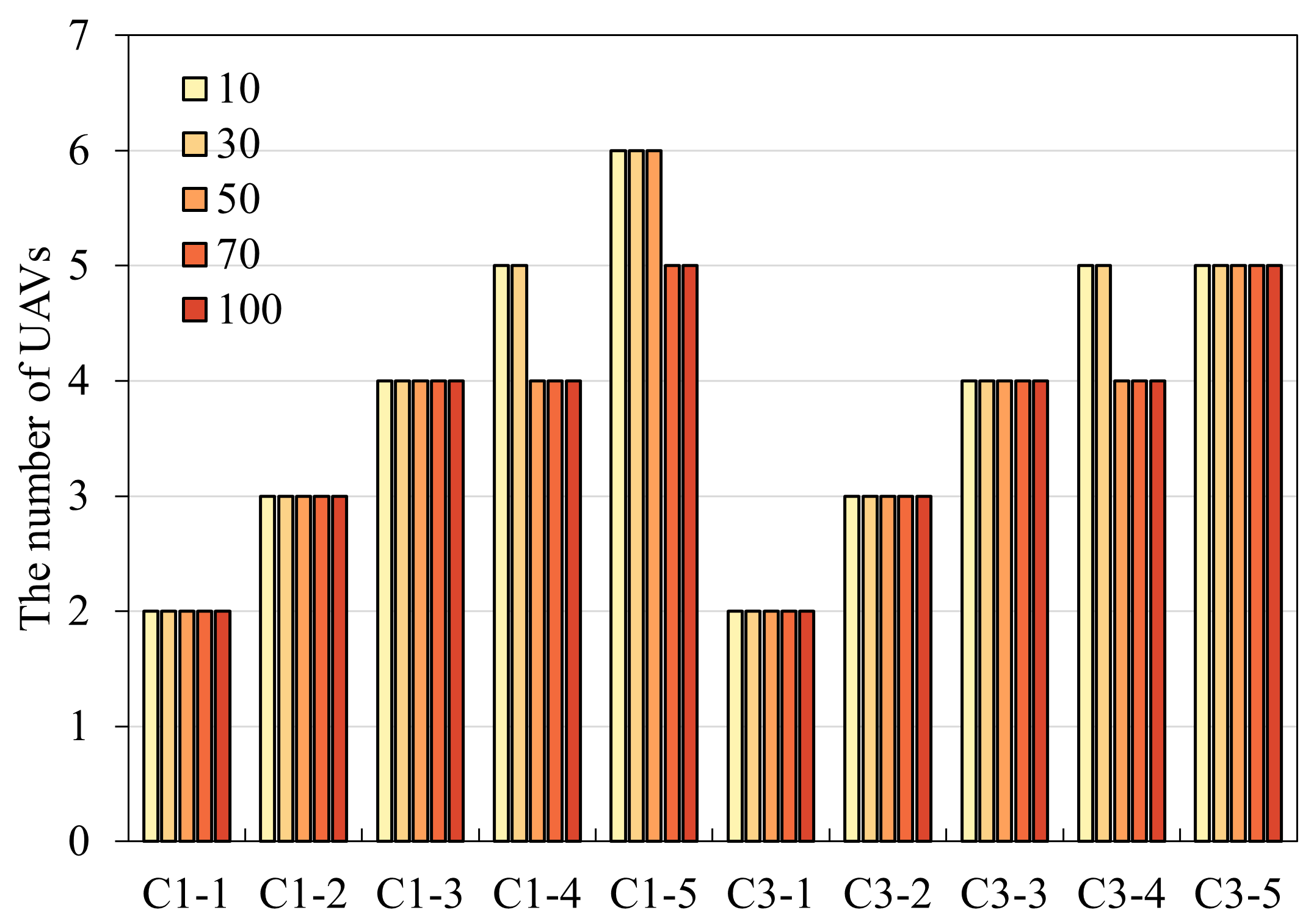}
				\caption{}\label{subFig5.10.1} 
			\end{minipage}
		\end{subfigure}
		
		\begin{subfigure}[t]{0.4\linewidth}
			\captionsetup{justification=centering} 
			\begin{minipage}[b]{1\linewidth}
				\includegraphics[width=1\linewidth]{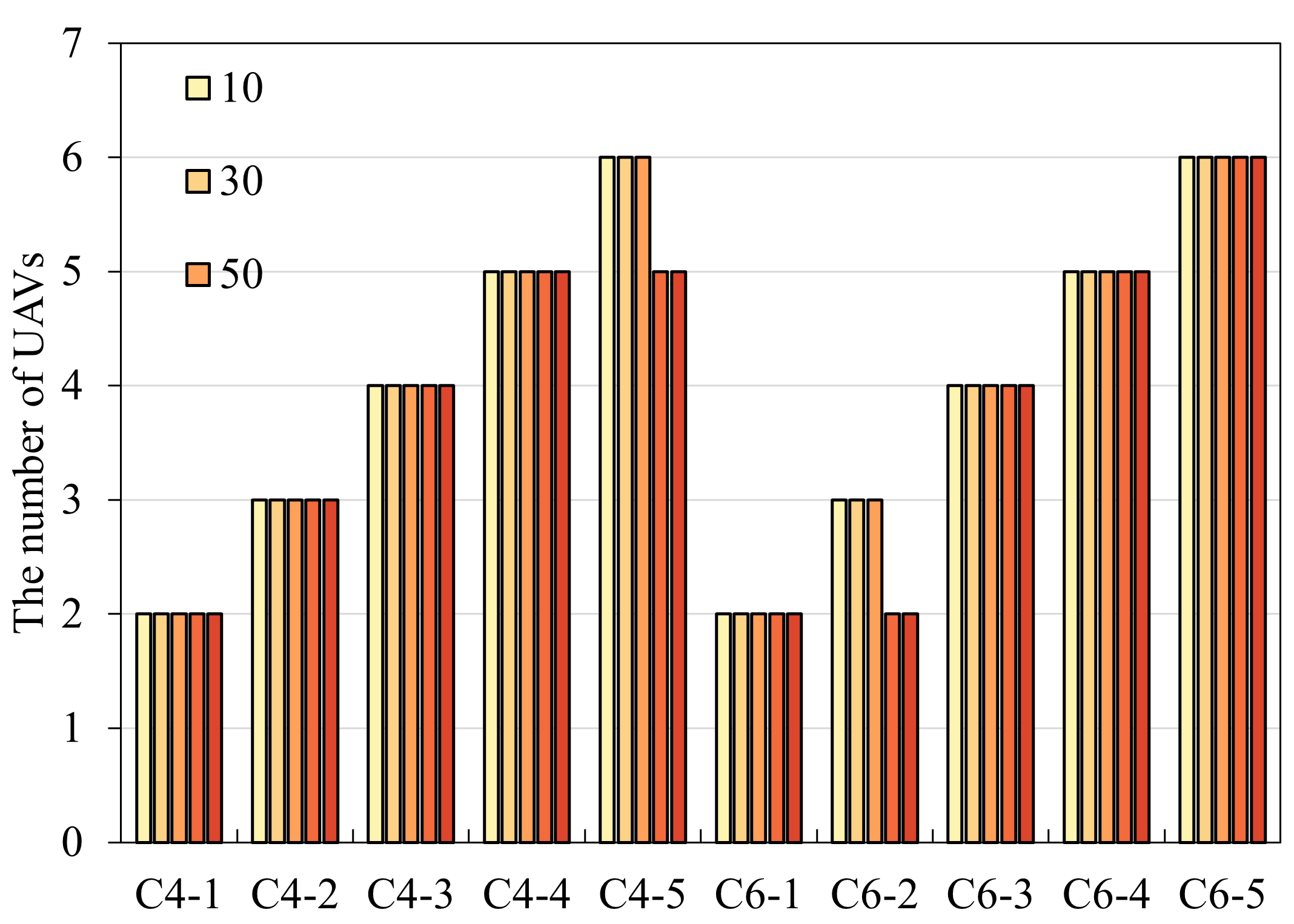}
				\caption{}\label{subFig5.10.2} 
			\end{minipage}
		\end{subfigure}\\	
	\end{tabular}
	\vspace{-2mm} 
	\caption{The number of used UAVs with different radius of disk neighborhoods}
	\label{fig11} 
\end{figure}

In this section, we evaluate the performance of AILS-VND-SOCP under different radius of disk neighborhoods (10, 30, 50, 70, 100). To this end, six instances are selected from both Type {I} and Type {II}, with each instance consisting of five sub-instances. AILS-VND-SOCP is executed 20 times on each instance, as summarized in Table A.7. Figs.~\ref{fig10}-\ref{fig11} illustrate the changes in flight distance and number of used UAVs and  as the radius of the disk neighborhoods varies.


Fig.~\ref{fig10} shows that as the radius of disk neighborhoods is increased, the flight distance decreases. In addition, for instances where the required nodes are relatively sparse, increasing the radius of the disk neighborhood often further reduces the number of used UAVs. For example, in instances C1-5 and C4-5 shown in Fig.~\ref{fig11}, the number of used UAVs is reduced by one when the disk neighborhood radius is set to 70 or 100. This suggests that expanding the disk neighborhood radius can contribute not only to shorter routes but also to fewer used UAVs.

\section{Conclusions}
\label{sec6}

In this study, we introduce CEMUAVGRP, in which a fleet of homogeneous UAVs conduct monitoring tasks containing nodes---each associated with a disk neighborhood (the coverage monitoring strategy)---and edges, aiming to minimize the total distance. The coverage monitoring strategy and edge routing significantly increase the complexity of the problem, making it particularly challenging to solve.

To address this problem, we propose a two-phase iterative method that decomposes the CEMUAVGRP into two phases. In the general routing phase, feasible routes are constructed for each UAV by considering the required nodes and edges, but without accounting for the disk neighborhoods. In the close-enough routing phase, representative points within the disk neighborhoods are optimized on the previously determined routes. Specifically, VND is employed in the general routing phase, while SOCP is used in the close-enough routing phase. These two phases are integrated within AILS framework and are executed iteratively until the predefined termination criteria are met. Extensive experiments on the CEMUAVGRP benchmark instances without disk neighborhoods \citep{campbell2023multi} demonstrate the strong optimization performance of AILS-VND-SOCP. For 150 benchmark instances, the best solutions obtained by AILS-VND-SOCP exhibit a gap of no more than 2.5\% compared to the best-known solutions. Notably, it outperforms the B\&C on 13 instances. Furthermore, the coverage monitoring strategy and re-increasing the acceptance threshold factor are shown to effectively reduce total flight distance.

In the future, this research can be extended in several directions. First, incorporating UAV recharging could further reduce monitoring costs and improve efficiency. This would introduce additional elements into the scenario, such as charging stations, and raise new challenges: when UAVs should recharge, which station they should choose, and which task should be prioritized after recharging. Second, required edges may also be split. Unlike disk neighborhoods of required nodes, splittable required edges have both direction and length, which influence the solution structure and substantially increase computational complexity.

\section*{Acknowledgments}
This work was supported in part by the National Natural
Science Foundation of China under Grant 62373380, in part by the China Scholarship Council under Grant 202306370294 and in part by the Natural Science Foundation of Hunan Province under Grant 2025JJ10007.

\bibliographystyle{cas-model2-names}

\bibliography{reference}


\end{document}